\documentclass{aa}  

\usepackage[english]{babel}           
\usepackage{graphicx,epsfig}
\usepackage{txfonts}
\usepackage{lscape}
\newcommand{\eg}{{\it{e.g.~}}}

\def \ie{{\rm i.e.}}
\def \eg{{\rm e.g.}}

\def\cc{\ifmmode{\,{\rm cm}^{-3}}\else{$\,{\rm cm}^{-3}$}\fi}
\def\cq{\ifmmode{\,{\rm cm}^{-2}}\else{$\,{\rm cm}^{-2}$}\fi}
\def\mic{\ifmmode{\,\mu{\rm m}}\else{$\mu$m}\fi}
\def\eccs{\ifmmode{\,{\rm erg}\,{\rm cm}^{-3} {\rm s}^{-1}}\else{$\,{\rm
erg}\,{\rm cm}^{-3} {\rm s}^{-1}$}\fi}
\def\ecqs{\ifmmode{\,{\rm erg}\,{\rm cm}^{-2}\,{\rm s}^{-1}\,{\rm
sr}^{-1}}\else{$\,{\rm erg}\,{\rm cm}^{-2}\,{\rm s}^{-1}\,{\rm sr}^{-1}$}\fi}
\def\deg{\ifmmode{^{\circ}}\else{$^{\circ}$}\fi} 
\def\pc{\ifmmode{\,{\rm pc}}\else{$\,{\rm pc}$}\fi} 
\def\kms{\ifmmode{\,{\rm km}\,{\rm s}^{-1}}\else{km s$^{-1}$}\fi} 
\def\kmspc{\ifmmode{\,{\rm km}\,{\rm s}^{-1}\,{\rm pc}^{-1}}\else{km
s$^{-1}$ pc$^{-1}$}\fi} 
\def\MJysr{\ifmmode{\,{\rm MJy\,sr}^{-1}}\else{$\,{\rm MJy\,sr}^{-1}$}\fi} 
\def\Kkms{\ifmmode{\,{\rm K\,km\,s}^{-1}}\else{$\,{\rm K\,km\,s}^{-1}$}\fi} 
\def\twCO{\ifmmode{\rm ^{12}CO}\else{$\rm^{12}CO$}\fi} 
\def\thCO{\ifmmode{\rm ^{13}CO}\else{$\rm^{13}CO$}\fi} 
\def \Cp{\ifmmode{\rm C^+}\else{$\rm C^+$}\fi} 
\def \CHp{\ifmmode{\rm CH^+}\else{$\rm CH^+$}\fi}
\def \CHtp{\ifmmode{\rm CH_2^+}\else{$\rm CH_2^+$}\fi} 
\def \CtH{\ifmmode{\rm C_2H}\else{$\rm C_2H$}\fi} 
\def\CHthp{\ifmmode{\rm CH_3^+}\else{$\rm CH_3^+$}\fi} 
\def \HCOp{\ifmmode{\rm HCO^+}\else{$\rm HCO^+$}\fi} 
\def \HtOp{\ifmmode{\rm H_3O^+}\else{$\rm H_3O^+$}\fi} 
\def \HCfiN{\ifmmode{\rm HC_5N}\else{$\rm HC_5N$}\fi} 
\def\wat{\ifmmode{\rm H_2O}\else{$\rm H_2O$}\fi} 
\def \oxy{\ifmmode{\rm O_2}\else{$\rm O_2$}\fi} 
\def \HH{\ifmmode{\rm H_2}\else{$\rm H_2$}\fi}
\def \Jone{\ifmmode{\rm {(J=1--0)}}\else{{(J=1--0)}}\fi} 
\def\Jtwo{\ifmmode{\rm {(J=2--1)}}\else{{(J=2--1)}}\fi} 
\def\Jthr{\ifmmode{\rm {(J=3--2)}}\else{{(J=3--2)}}\fi} 
\def\Jfou{\ifmmode{\rm {(J=4--3)}}\else{{(J=4--3)}}\fi} 
\def\Jfiv{\ifmmode{\rm {J=4--3}}\else{{J=4--3}}\fi} 
\def \Ta{\ifmmode{\rm T_A}\else{$\rm T_A$}\fi} 
\def \Tas{\ifmmode{\rm T_A^*}\else{$\rm T_A^*$}\fi} 
\def \Tmb{\ifmmode{\rm T_{mb}}\else{$\rm T_{mb}$}\fi} 
\def \Tr{\ifmmode{\rm T_r}\else{$\rm T_r$}\fi} 
\def \Trs{\ifmmode{\rm T_r^*}\else{$\rm T_r^*$}\fi}

\begin{document}

\title{Models of turbulent dissipation regions in the diffuse interstellar medium} 

\author{\vspace{2cm}
  B. Godard \inst{1}, 
  E. Falgarone \inst{1} 
  \and 
  G. Pineau des For\^ets \inst{2,1}
}

\institute{LRA/LERMA, CNRS UMR 8112, \'Ecole Normale Sup\'erieure 
\& Observatoire de Paris, Paris
  \and
  Institut d'Astrophysique Spatiale, CNRS UMR 8617, Universit\'e Paris-Sud, Orsay}

 \date{Received 14 august 2008 / Accepted 11 december 2008}

\abstract{}
{Supersonic turbulence is a large reservoir of suprathermal energy in
the interstellar medium. Its dissipation, because it is intermittent in space
and time, can deeply modify the chemistry of the 
gas. This is clearly seen in the framework of shock chemistry. Intense
turbulent dissipation also occurs in regions of large 
velocity shears, sharing with shocks the property of 
intermittency. Whether these bursts of dissipation, short-lived and localized,
have a measurable impact on molecular abundances in the diffuse medium,
and how the chemical enrichment they drive compares to observations, 
are the questions we address here.}
{We further explore a hybrid method 
to compute the chemical and thermal evolution of a magnetized
dissipative structure, under the energetic constraints provided by the
observed properties of turbulence in the cold neutral medium.  For the
first time, we model a random line of sight by taking into account the
relative duration of the bursts with respect to the thermal and
chemical relaxation timescales of the gas.  The key parameter is the
turbulent rate of strain $a$ due to the ambient
turbulence. With the gas density, it controls the size of the
dissipative structures, therefore the strength of the burst. 
It also sets the relative importance of viscous dissipation and ion-neutral
friction in the gas heating and chemical enrichment.
}
{For a large range of
rates of strain and densities, the models of turbulent dissipation 
regions (TDR) reproduce the \CHp\ column densities
observed in the diffuse medium and their correlation with highly
excited \HH.  They do so without producing an excess of CH. 
As a natural consequence, they reproduce
the abundance ratios of \HCOp/OH and \HCOp/\wat,
and their dynamic range of about one order of magnitude  observed in diffuse
gas. Large C$_2$H and CO abundances, also related to 
those of \HCOp, are another outcome of the TDR models 
that compare well with observed values.
Neutral carbon exceeds the abundance expected at
ionization equilibrium, in agreement with fine-structure line
observations.  The abundances and column densities computed for CN, HCN and
HNC are one order of magnitude above PDR model predictions, although
still significantly smaller than observed values.
The dependence of our results on the rate of strain and density reveals
that the chemical enhancements are in better agreement with observations
if the dissipation is dominated by  ion-neutral friction, involving
shear structures of thickness $\sim 100$ AU.
}
{}


   \authorrunning{B. Godard et al.}
   \titlerunning{Models of turbulent dissipation regions}
   \maketitle
%

\section{Introduction}
The diffuse medium has a major contribution to the mass
of the interstellar medium (ISM) in galaxies like the \object{Milky Way} and as
such is a key player in the star formation process. Although it is the
first component of the ISM to have been discovered, and later on
extensively analyzed through absorption measurements of atoms, ions
and molecules (see the review of Snow \& McCall, 2006),
its structure and properties remain a challenge in
many respects: \\
(1) long thought to consist of two stable phases - the
warm and cold neutral medium (WNM at temperatures $T \sim$ 8000 K and
CNM at $T \sim$ 100 K) in thermal pressure
equilibrium - a significant fraction of its emission is now detected at
temperatures covering the whole range between those of the CNM and WNM
(Heiles \& Troland 2003),\\ 
(2) the CNM is turbulent with supersonic velocities, yet the velocity
and density power spectra carry the signature of the Kolmogorov
power spectrum for incompressible turbulence (Miville-Desch\^enes
et al. 2003),\\
(3) its spatial and velocity structure is even a greater
challenge since it has to reconcile the existence of structures
observed at all scales in emission and a remarkable similarity of line
profiles observed in absorption (see the discussion in Liszt \&
Lucas 1998 who question the elusive dimensionality of the diffuse
medium).  Since many tracers of the kinematics and small-scale
structure of the diffuse ISM are molecular lines, 
clues are likely to be found in its
surprisingly rich, but poorly understood, chemistry.

One major puzzle in this chemistry, raised by the detection of \CHp\
in almost every line of sight sampling the CNM, persisted for decades
because no formation path was found to be efficient enough in the
diffuse ISM (Black, Dalgarno \& Oppenheimer 1975, Black \& Dalgarno
1977).  It is rooted in the fact that in such diffuse gas, \CHp\ forms
via a highly endoenergetic reaction \Cp\ + \HH\ ($\Delta E/k$ = 4640
K) unlikely to proceed at the low temperatures of the CNM. Similarly,
one way to activate the oxygen chemistry in the diffuse medium, and
therefore the formation of OH and \wat, involves the reaction of O +
\HH\ which has an energy barrier of 2980 K.

A possibly related issue is the existence of rotationally
excited H$_2$ in the diffuse gas. FUSE observations have revealed
large populations in the $J>2$ levels of H$_2$, inconsistent with
fluorescence generated by the ambient UV field (Sonnentrucker et
al. 2003; Gry et al. 2002; Nehm\'e et al. 2008a,b; Martin-Za\"\i di
et al. 2005; Lacour et al. 2005).  This excited H$_2$ could be
localized in circumstellar material\footnote{ Here, we deliberately
overlook most of the early Copernicus results on \HH\ high-$J$ lines
obtained in the direction of hot stars (\eg\ Spitzer et al. 1973;
Savage et al. 1977; Shull \& Beckwith 1982) that led to the conclusion
that excited \HH\ absorption is occurring in circumstellar material
heated/shocked by the star itself.}, but it also has been detected in
the direction of late B stars, devoid of circumstellar matter, as in
the data of Gry et al. (2002). Stellar UV photons are therefore
unlikely to contribute significantly to the UV-pumping of H$_2$.  In
particular Lacour et al. (2005) find an increase of the Doppler
parameter of the H$_2$ lines with $J$, supporting the existence of a
warm component that cannot be heated by UV photons.  They argue that
this warm component cannot be due to H$_2$ formation pumping, as
proposed by Sternberg \& Dalgarno (1995) in dense PDRs, because it
would require an H$_2$ formation rate larger than that inferred from
observations, and would not reproduce the observed column densities of
CH$^+$ found to correlate with excited H$_2$ (Spitzer et
al. 1974; Snow 1976, 1977; Frisch \& Jura 1980; Lambert \& Danks
1986).
				   
ISO-SWS observations further support the possible existence of a
 small fraction of warm gas in the Galactic diffuse medium by
 revealing its pure rotational line emission (Falgarone et
 al. 2005). Interestingly, the ratio $N({\rm H}_2)_{warm}/N_{\rm H}
 \sim 2 \times 10^{-4}$, where $N({\rm H}_2)_{warm}$ is the \HH\
 column density in levels $J\geq 3$, is the same across the
 Galactic diffuse medium as in the direction of nearby late B stars.
 Recent {\it Spitzer} observations have confirmed the ISO-SWS line
 flux values (Verstraete et al. in preparation).

Both the observed abundances of CH$^+$ and column densities of
rotationally excited H$_2$ suggest that large amounts of suprathermal
energy are deposited in the cold diffuse medium.  One obvious
reservoir of suprathermal energy in the ISM is its turbulent kinetic
energy. Attempts at incorporating this energy in the chemical networks
of magneto-hydrodynamical (MHD) shocks have been partly successful at
reproducing the observed properties of the diffuse medium (Pineau des
For\^ets et al. 1986, Draine \& Katz 1986, Flower \& Pineau des
For\^ets 1998). Other routes have been explored, involving
dynamic interactions of the gas and the star cluster in the \object{Pleiades}
(White 1984, 2003; Ritchey et al. 2006), turbulent transport between
the WNM and CNM (Lesaffre et al. 2007) and turbulent dissipation
taking place in regions of large velocity shears.  Turbulence being
intermittent in space and time (see the review of Anselmet, Antonia \&
Danaila 2001), velocity shears may locally be large enough to drive
large local heating rates and trigger the endoenergetic reactions of
carbon and oxygen chemistries in the diffuse ISM (Falgarone,
Pineau des For\^ets \& Roueff 1995). Along these lines, Joulain et
al. (1998, hereafter J98) have explored the role of ion-neutral
decoupling induced, in the weakly ionized diffuse medium, by the sharp
gas accelerations in the regions of largest velocity-shear and its
impact on ion-neutral chemistry, in particular the formation of CH$^+$
and HCO$^+$.  Falgarone et al. (2006) have analysed the thermal and
chemical relaxation phase in the evolution of a gas chemically
enriched in a dissipation burst.

The observational data probing the molecular diversity and richness of
the diffuse ISM are not restricted to CH$^+$ and HCO$^+$.  In their
long-lasting effort dedicated to unravelling molecular abundances in
the diffuse ISM, Liszt \& Lucas (2002 and references therein) have
provided us with invaluable constraints. Not only did they show that the
abundances of several molecules stay proportional to each other, with
very well defined abundance ratios, but they found that the column
densities of these molecules vary by more than one order of magnitude
across clouds that all have about the same total hydrogen column
density, corresponding to diffuse and translucent clouds. They also
revealed the remarkable similarity of \HCOp\ and OH line profiles,
all the more surprising for an ion and a neutral species, differently coupled
to the magnetic field.  On the contrary, the high-spectral resolution
spectra of Crane, Lambert and Sheffer (1995) convincingly showed that
the \CHp\ line profiles are definitely broader and less Gaussian than
those of CH, along the same lines of sight, while Lambert, Sheffer \&
Crane (1990) found that, in the direction of \object{$\zeta$ Oph}, the CH
line profiles could be seen as the sum of a broad component similar to
the \CHp\ profile, and a narrow one, close to that of the CN
line. A similar result was obtained by Pan et al. (2004, 2005)
towards stars of the \object{CepOB2} and \object{CepOB3} regions.
These sets of results suggest that the velocity field is involved in the
origin and the evolution of these molecules, and does so differently for each
species.

It is therefore challenging to compare these available
observations with models of a random line of sight across the
diffuse ISM, where active dissipation bursts coexist with others in
their relaxation phase. In particular, the possibility that a number
of transient events may dominate the observed molecular column
densities has never been addressed.  This is what we do in the present
paper.  We restrict our study to densities lower than $n_{\rm
H}= 200$ \cc\ because, as will be seen, at higher densities and for
the turbulent energy observed in the CNM, turbulent dissipation does
not heat the gas enough to open the endoenergetic barriers
mentioned above. We cannot rule out, though, rare dissipation bursts 
of exceptional intensity that would be able to heat still denser gas 
to the required temperature. We extend the previous studies in a way that
allows us to explore the parameter domain, in particular those characterizing
the ambient turbulence. We also extend the chemical network.  We model
a random line of sight across the diffuse medium and compare the
predicted column densities of a variety of molecular species to the
observations.  The dynamic steady state is computed and described in
Section 2, the chemistry in the active dissipation and relaxation
phases is presented in Sections 3 and 4.  The modelling of a line of
sight is discussed in Section 5. Comparisons of computed column
densities with observations, as well as excitation diagrams of \HH,
are shown in Section 6 and these results are discussed in
Sect. \ref{Discussion}.

\section{Steady state of a magnetized vortex in a weakly-ionized diffuse gas} \label{Dynamics}

\subsection{The neutral flow} \label{Hydroprop}

Turbulence in the diffuse ISM is supersonic with respect to its cold phase,
the CNM.  Supersonic turbulence dissipates in shocks and regions of
large velocity shear (Kritsuk et al. 2007). 
Their respective importance has been studied in numerical simulations.
Porter et al. (1992; 1994) and Pavlovski, Smith \& Mac Low (2006) 
showed that most of the turbulent 
kinetic energy is rapidly transferred to high wavenumber 
non-compressible modes,
once the shocks generated by supersonic turbulence have started to interact,  
reducing the role of compressible modes (shocks) in turbulent dissipation. 
In the so-called quiescent ISM (i.e. far from star forming regions), 
the smooth observed lineshapes support the view of a
turbulence devoid of strong shocks (Falgarone et al. 1994) implying that
the dissipation preferentially occurs in shear-layers.

Dissipative structures are modelled 
as shear-layers belonging to a solution of the
Helmholtz equation for vorticity, close to the Burgers vortex adopted
in J98: this analytical solution has the merit that it has only two
free parameters that describe the balance between the stretching action of
the large scales and the diffusion of vorticity across the vortex
edge, at small scale. It provides an analytical framework in which we can
compute the effect of partial decoupling between ions and neutrals
upon the steady state configuration of velocity and magnetic field and thus 
explore the effect of both the ion-neutral friction and viscous heating
upon the chemical network.

The modified Burgers vortex is an axisymmetric
solution elaborated in atmospheric
sciences by Nolan \& Farrell (1998). It is identical to the Burgers vortex 
at small radii $r$, in cylindrical
coordinates $\left( r, \theta, z \right)$, and differs from it at large radii
in the sense that the radial inflow velocity  
is not divergent at infinity:
\begin{equation}\label{EqUr}
u_r(r) = -\frac{a}{2} r \cdot e^{-\beta r^{2}}
\end{equation}
where $a$ is the turbulent rate of strain (in s$^{-1}$) and $\beta$
 (in cm$^{-2}$) describes the cut-off of the radial velocity. The
 axial velocity $u_z$, the vorticity $\omega_z$  
and the orthoradial
 velocity $u_{\theta}$ 
are inferred from the Helmholtz and continuity equations:

\begin{equation}\label{EqUz}
u_z(r) = a z \cdot e^{-\beta r^{2}} \cdot \left( 1 - \beta r^{2} \right),
\end{equation}
\begin{equation}\label{EqOm}
\omega_z (r) = \omega_0 \cdot e^{-\frac{a}{4\nu\beta}
\left[1-e^{-\beta r^{2}} \right] },
\end{equation}
\begin{equation}\label{EqUtheta}
u_{\theta}(r) = \frac{1}{r} \int_{0}^{r} r' \omega_z(r') dr'
\end{equation}
where $\omega_0$ is the peak of vorticity and $\nu$ is 
the kinematic viscosity. Any vortex is therefore
entirely defined by three parameters, $a$, $\beta$ and $\omega_0$.

Note that, according to the radial dependence of the vorticity, the
same equilibrium vortex radius $r_0$
 as for the Burgers vortex can be defined,
\begin{equation} \label{radiusr0}
r_0^2= 4\nu/a
\end{equation}
involving the two quantities that act on it,
the rate of strain $a$ and the viscosity $\nu$. 
Accordingly, the vortex crossing time 
\begin{equation}
\tau_c= \int_{r}^{k r} \frac{dr'}{u_r(r')}=\frac{2}{a} \ln (1/k)
\end{equation}
for any constant $k<1$ depends only on the rate of strain, while the vortex
period, defined as
\begin{equation}
P = \frac{r_0}{u_{\theta}(r_0)}
\end{equation}
depends on $\nu$, $a$, and $\omega_0$.

Because vorticity is radially non uniform there is a differential rotation of
the fluid within the structure which induces a viscous dissipation rate:
\begin{equation}\label{EqGam1}
\Gamma_{nn} = \sum_{i,j} \pi_{ij} \frac{\partial{u_i}}{\partial{x_j}}
\end{equation}
where $\pi_{ij}$ is the stress tensor. This rate is written in
cylindrical coordinates:
\begin{equation}\label{EqGam2}
\frac{\Gamma_{nn}}{\eta} = 
\left[\frac{\partial{u_{\theta}}}{\partial{r}} -
\frac{u_{\theta}}{r}\right]^{2} +
\left(\frac{\partial{u_z}}{\partial{r}}\right)^{2} +
2\left(\frac{\partial{u_r}}{\partial{r}}\right)^{2} +
2\left(\frac{u_r}{r}\right)^{2} +
2\left(\frac{\partial{u_z}}{\partial{z}}\right)^{2}
\end{equation}
where $\eta=\rho \nu$ is the dynamic viscosity in a gas 
of density  $\rho$. It is computed  for  
hydrogen atoms using the Kay \& Laby (1966) tables of physical and 
chemical constants
: $\eta = 6 \times 10^{-6}\, T_k^{1/2}$ g cm$^{-1}$ s$^{-1}$ where
$T_k$ is the gas kinetic temperature.

In all the following, it is assumed that the fluid description of the 
gas motions is justified because the mean free path of H atoms
$\lambda_{\rm H-H} = 0.23(n_{\rm H}/50$ cm$^{-3} )^{-1}$ AU in the diffuse
medium, for a H-H elastic collision cross section\footnote{ For comparison
    the H$_2$-H$_2$ elastic collision cross section is
    $\sigma_{\rm{H}_2-\rm{H}_2} \sim 3 \,\, 10^{-15} \rm{cm}^{2}$ (Monchick et
    al. 1980).} $\sigma_{\rm H-H} = 5.7 \times 10^{-15}$ cm$^{2}$ (Spitzer
  1978), is smaller than all the lengthscales involved in the model.

\subsection{Interstellar constraints on the vortex parameters} \label{Constraints}

As said above, each vortex is defined by a set of three independent 
parameters, $a$, $\beta$ and $\omega_0$. 
The cut-off parameter $\beta$ has 
no influence either on the dynamics or on the chemistry
of the structure as long as $u_r$ is small compared to
$u_{\theta,max}$. Hereafter, $\beta$ is chosen in order to satisfy this
condition. In other words, turbulent dissipation in
the vortex happens through radial vorticity distribution, not through
a radial flux of matter. In these conditions, the dominant
contribution to the viscous dissipation is the first term in
Eq. (\ref{EqGam2}).

Numerical simulations of incompressible turbulence (Jimenez
1997) and experiments (Belin et al. 1996) have shown that the maximum
tangential velocity of  filaments of vorticity $u_{\theta,max} \sim \omega_0
r_0$ is of the order of the rms velocity dispersion of the ambient turbulence
$\sigma_{turb}$. Since the equilibrium radius 
$r_0$ is set by the rate of strain $a$ (Eq. \ref{radiusr0}), 
$\omega_0$ is also determined.
In the case of interstellar turbulence, it implies that
the orthoradial velocity in the vortex is supersonic with respect to the 
cold gas. It is noteworthy that 
slightly supersonic Burgers 
vortices have been found in experiments of rotating magnetized plasmas by 
Nagaoka et al. (2002) in conjunction with an inner density hole. Moreover, 
as will be seen later, the gas being violently and rapidly heated 
in the layers of largest orthoradial velocity, the Mach number there
is likely to drop below unity. 
The only free parameter left in the vortex description is therefore the
rate of strain $a$, although the gas density is in fact a free parameter that
determines the vortex size, through the kinematic viscosity
$\nu$ (Eq. \ref{radiusr0}).  

\subsection{Magnetic field configuration and ionized flow} \label{Magnetprop}

The configuration of the magnetic field and ions reached once the
above vortex has developed in the partially ionized gas is numerically
computed.
The ions are \emph{initially} at rest, threaded by a uniform magnetic field
parallel to the $z$ axis, $\mathbf{B_0} = B_{0} \mathbf{k}$. The ions
are predominantly C$^{+}$, the neutrals are mainly composed of H and
H$_2$ and the ionization degree, $x = 2 \times 10^{-4}$, is weak\footnote{
  Computed for a diffuse molecular cloud of density $n_{\rm H} \sim
50$ cm$^{-3}$, temperature $T_k \sim 100$ K, illuminated by the
standard interstellar radiation field (ISRF).}.  At $t=0$, the ions are
suddenly put into motion by friction with the vortex that developed in
the neutral gas \ie\ the three components of the neutral gas velocity
$\mathbf{u_n}$ are those of the vortex given by Eqs. (1),
(2) and (4).  Boundary conditions are provided by the assumption that
the vortex has a finite length $L_V$, apodised over a length $C_V$.

The  alignment of $\omega$ with the ambient magnetic field
 is supported by the results of numerical simulations.
 Brandenburg et al. (1996) showed that in MHD turbulence, magnetic
 field and vorticity vectors tend to align with each other. More
 recently Mininni et al. (2006a,b)
 observed a similar behaviour in their 1536$^{3}$ numerical 
 simulations. 

Under these assumptions, we compute the two-dimensional
time-dependent evolution of the ion velocity $\mathbf{u_i}$ and the
magnetic field $\mathbf{B}$. We neglect the retro-action of the ions
upon the neutral motions because, for densities in the range 10-200
cm$^{-3}$ and an ion-neutral drift velocity comparable to $u_n$ (see
Fig. 1),  the friction force $F_{in}$ they
exert on the neutrals is negligible compared to the advection force in
the vortical motion: $F_{in} \sim 10^{-3} (l/10 {\rm AU}) 
\times \rho_n \, u_n. \nabla u_n $, $l$ being the spatial scale for the 
variation of $u_n$, in the range 10 to 100 AU.  
The neutrals velocity components are
therefore those of the vortex (Eqs. (1), (2) and (4)) at any time.

In the interstellar medium $\mathbf{B}$ is
frozen in the charged fluid (Spitzer 1978) and its evolution 
is simply written:
\begin{equation} \label{EqFrozen}
\frac{\partial \mathbf{B}}{\partial{t}} + \nabla \times \left( \mathbf{B}
\times \mathbf{u_i} \right) = 0.
\end{equation}

Neglecting the pressure gradients in the evolution equation of 
the ionized flow (this assumption is justified in Sect. \ref{Discussion})
leads to:
\begin{equation} \label{EqIons}
\frac{\partial \mathbf{u_i}}{\partial{t}} + \left(\mathbf{u_i} \cdot \nabla
\right) \mathbf{u_i} = \frac{\left<\sigma v\right>_{in}}{\left(\mu_n+\mu_i\right)}
 \rho_n \left(\mathbf{u_n}-\mathbf{u_i} \right) +
\frac{1}{4 \pi \rho_i} \left(\nabla \times \mathbf{B} \right) \times \mathbf{B}
\end{equation}
where $\mu_n$ and
$\mu_i$ are the mean mass per particle of the neutrals (H, H$_2$) and ions
(mostly C$^{+}$) respectively.
$\left<\sigma v\right>_{in}=2.2 \times 10^{-9}$cm$^{3}$ s$^{-1}$ 
is the momentum
transfer rate coefficient between the ionized and neutral fluids 
calculated by Flower \& Pineau des For\^ets (1995, Appendix A), and 
close to the Langevin rate.

\begin{figure}[!h]
\begin{center}
\includegraphics[width=9cm,angle=0]{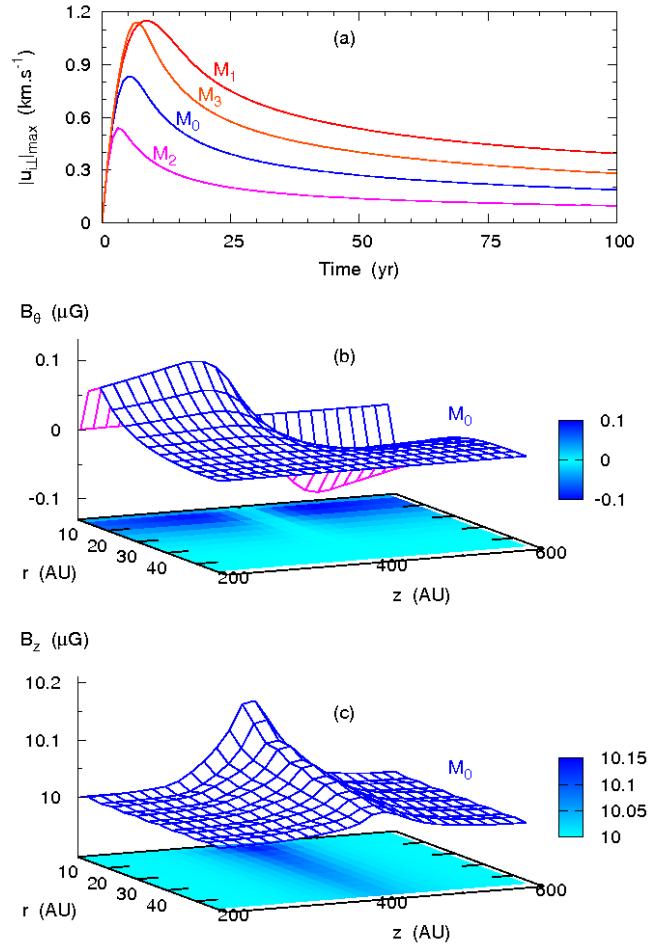}
\caption{Magnetic properties of the vortex. Panel (a):
  evolution of $\left| u_{i\perp} \right|_{max}$, the maximum ion velocity
  component perpendicular to the axis $z$, as a function of time. The
  different curves correspond to the models $M_0$, $M_1$, $M_2$ and $M_3$,
  (see Table \ref{TabMagnet}). Panels (b) and (c): orthoradial and axial
  components of the magnetic field at $t = 100$ yr for the model
  $M_0$ as functions of $r$ and $z$.}
\label{FigMagnet}
\end{center}
\end{figure}

\begin{table}[!h]
\begin{center}
\begin{tabular}{l l l l l l l}
\hline
     &    & & $M_0$ & $M_1$ & $M_2$ & $M_3$ \\
\hline
Magnetic field & $B$ & $\mu$G & 10    & 5     & 10    & 10    \\
Vortex length  & $L_V$ &  AU     & 200   & 200   & 100   & 200   \\
Apodisation length & $C_V$ & AU     & 100   & 100   & 50    & 50    \\
\hline
\end{tabular}
\caption{Parameters of the four models of Fig. \ref{FigMagnet}. The
  density $n_{\rm H}$ and the turbulent rate of strain are fixed: $n_{\rm H} = 50$
  cm$^{-3}$, $a = 5 \times 10^{-10}$ s$^{-1}$.}
\label{TabMagnet}
\end{center}
\end{table}

We integrate Eq. (\ref{EqFrozen}) and Eq. (\ref{EqIons}) by means of a
two dimensional implicit scheme using the Alternating Direction Implicit
method (\emph{ADI}). To validate our approach we also use two other integration
schemes: an explicit and an implicit without the \emph{ADI} method. The results
of our $300\times200$ points grid simulations are displayed in
Fig. \ref{FigMagnet}. Panel (a) shows the evolution of
$\left| u_{i\perp} \right|_{max}$, the maximum ion velocity component
perpendicular to the axis $z$, as a function of time. The curves correspond
to the four models presented in Table \ref{TabMagnet}. Panels (b) and
(c) show the orthoradial and axial components of $\mathbf{B}$ for the model
$M_0$ at $t=100$ yr.

Fig. \ref{FigMagnet}a shows that the ions, initially at rest, are put
into motion by the friction force from the neutrals in the vortex. This
motion (including its boundary conditions) generates an orthoradial
component $B_{\theta}$ of $\mathbf{B}$ and a gradient of $B_z$. These
terms in turn induce magnetic tension and pressure gradient
(see Eq. \ref{EqFrozen}), two forces resisting the orthoradial
entrainment exerted by the neutrals. 
After  $\sim 100$ yr,
$\left|\mathbf{u_{i\perp}}\right|_{max} < 0.4$ km s$^{-1}$ which is
small compared to $\left|\mathbf{u_{n\perp}}\right|$ in the vortex
(see Fig. \ref{FigDyn} in Sect. \ref{ThermVA}) for all models. A 
steady state is reached in which the ions are almost back to rest and
the magnetic field slightly helical (Fig. \ref{FigMagnet}b).
A large and steady state
drift is set between the ion and neutral orthoradial velocities with an
amplitude close to the orthoradial velocity of the neutrals in the
vortex.

Such a drift has a deep impact on 
the chemistry of the gas, as was shown by J98, and contributes 
to the dissipation of its turbulent energy. The additional heating term
due to the ion-neutral friction is written:
\begin{equation}\label{EqGamin}
\Gamma_{in} = \frac{\rho_n \rho_i}{\mu_n+\mu_i} \left<\sigma v\right>_{in}
\frac{\mu_i}{\mu_n+\mu_i} \left( \mathbf{u_i}-\mathbf{u_n} \right)^{2}.
\end{equation}

We also found that there is only a very small ion-neutral drift in the $z$
direction because the magnetic field is only slightly helical.
Since in addition the contributions of the spatial variations of $u_z$ to the
heating term $\Gamma_{nn}$ are negligible (Eq. \ref{EqGam2}), we consider the
vortex as invariant along the axis $z$ and restrict our study of the spatial
and time dependences to  those  occurring radially.

In the next sections we focus on the rapid thermal and 
chemical evolution of the gas trapped in such a steady state structure, 
and follow its thermal and chemical relaxation, once the vortex has blown-up.

\section{The active phase}\label{VortexActif}

\subsection{Numerical modelling} \label{NumVortexActif}

As in J98, we follow the Lagrangian evolution of a fluid particle
trapped in the steady state vortex configuration.  Because the vortex
crossing time $\tau_c$ (see Sect. \ref{Hydroprop}) is comparable to
the chemical timescales, we compute non-equilibrium chemistry coupled
to the time-dependent thermal evolution.  
The initial gas temperatures and molecular abundances are those 
of a steady state diffuse cloud of density $n_{\rm H}$, illuminated by 
the ambient interstellar radiation field (ISRF) (Draine 1978)  
scaled by the factor $\chi$, and shielded by the extinction 
$A_V$. The cosmic ray ionization rate $\zeta$ and the
elemental abundances are given in Table \ref{TabDiffPar}.

The neutrals and the ions are treated separately\footnote{
The ions and the electrons are treated as a unique fluid at a
temperature $T_i$ because: (1) the ion-electron velocity drift induced by the
magnetic field fluctuations in the model is $\sim 1$ cm
s$^{-1}$, negligible compared to the thermal velocities, (2) the ion-electron
temperature equipartition time is $\sim 0.1 \,\, (T/1000\, \rm{K})^{3/2}
( \rm{\emph{$n_i$}} /3\,10^{-3}\, \rm{cm}^{-3})^{-1}$ yr (Spitzer 1978), 
also negligible compared to the dynamic timescales.
} 
and a fluid particle
is defined at each time by its position $r$, the neutral and ionized
velocity fields $\mathbf{u_n}$ and $\mathbf{u_i}$, the mass densities
$\rho_n$ and $\rho_i$, the temperatures of the neutrals $T_n$ and ions
$T_i$, and the abundances $n(X)$ of each
species. The system therefore comprises 11 dynamic time-dependent
variables ($r$, $n_i$, $n_n$, $\rho_i$, $\rho_n$, $T_i$, $T_n$,
$u_{nr}$, $u_{n\theta}$, $u_{ir}$, $u_{i\theta}$).  Our chemical
network originates from the Meudon PDR code (Le Petit et al. 2006). It
incorporates 90 species interacting through 1524 reactions. Those
include the formation of H$_2$ on dust, the photoprocesses and the
processes induced by the cosmic rays. We also compute the
time-dependent evolution of the populations of the 18 first
ro-vibrational levels of H$_2$ (corresponding to $T_{ex} < 10^4$ K).

\begin{table}[!h]
\begin{center}
\begin{tabular}{l l l l c}
\hline
Density                & $n_{\rm H}$  & cm$^{-3}$ & 10 - 200  &  \\
Radiation field        & $\chi$ &           &  1  &  \\
Extinction             & $A_V$  & mag       & 0.1 - 1   &   \\
Cosmic ray ionization rate & $\zeta$ & s$^{-1}$ & $3 \times 10^{-17}$ & a \\ 
\hline
Elemental abundances  & & & $\left[\textrm{X}\right]/\left[\textrm{H}\right]$ &\\
Helium                 & $\left[\textrm{He}\right]$ & & $1.00 \times 10^{-1}$ & \\
Carbon                 & $\left[\textrm{C}\right]$  & & $1.38 \times 10^{-4}$ & \\
Oxygen                 & $\left[\textrm{O}\right]$  & & $3.02 \times 10^{-4}$ & \\
Nitrogen               & $\left[\textrm{N}\right]$  & & $7.94 \times 10^{-5}$ & \\
Sulfur                 & $\left[\textrm{S}\right]$  & & $1.86 \times 10^{-5}$ & \\
Iron                   & $\left[\textrm{Fe}\right]$ & & $1.50 \times 10^{-8}$ & \\
\hline
\end{tabular}
\caption{Physical conditions and elemental abundances of the gas in the
  TDR models. a - Dalgarno (2006).}
\label{TabDiffPar}
\end{center}
\end{table}

The system of variables is therefore driven by a set of
119 first-order coupled differential equations that are integrated along the
fluid particle trajectory. To ensure that the time step is consistent with
the variations of all dependent variables we use the DVODE differential
equation solver (Brown et al. 1989).

The evolution of the thermal energy densities $U_n$ and $U_i$ is given by:
\begin{equation} \label{EqUn}
\frac{dU_n}{dt} = \frac{d}{dt}\left(\frac{3}{2}n_nkT_n\right) = 
 B_n + \Gamma_{nn} + \Gamma_{in}+ \Gamma_{en}
\end{equation}
\begin{equation} \label{EqUi}
\frac{dU_i}{dt} = \frac{d}{dt}\left(\frac{3}{2}n_ikT_i\right) =  
B_i + B_e + \frac{\mu_n}{\mu_i} \Gamma_{in} + \frac{\mu_n}{\mu_e} \Gamma_{en}
\end{equation}
where $B_n$, $B_i$ and $B_e$ are the sums of all the heating and cooling
rates of the neutrals, the ions and the electrons respectively, not induced by
turbulent dissipation, and $\Gamma_{nn}$ and $\Gamma_{in}$ the
heating rates induced by turbulent dissipation defined in the previous section.
$\Gamma_{en}$ is the heating rate due to the electron-neutral drift, a term
taken into account in the code but negligible compared to $\Gamma_{nn}$ and
$\Gamma_{in}$.
The cooling rates include the radiative de-excitation of the
ro-vibrational levels of H$_2$, of the fine structure levels of C$^{+}$, C and
O and of the rotational levels of H$_2$O, OH and CO.

Lastly, since ions and neutrals are decoupled, we use the approximation of
Flower et al. (1985) for the calculation of the chemical rate
coefficients. The cross-section of a 2-species reaction is integrated over a
Maxwellian velocity distribution at an effective temperature
\begin{equation} \label{EqTeff}
T_{eff}  = \frac{m_1T_2+m_2T_1}{m_1+m_2} + \frac{1}{3k}
\frac{m_1m_2}{m_1+m_2} u_D^{2} 
\end{equation}
where $m_1$, $m_2$, $T_1$ and $T_2$ are respectively the masses and the
temperatures of the 2 reactants and $u_D$ their relative drift velocity.

\subsection{Thermal evolution of the gas} \label{ThermVA}

\begin{figure}[!h]
\begin{center}
\includegraphics[width=9cm,angle=0]{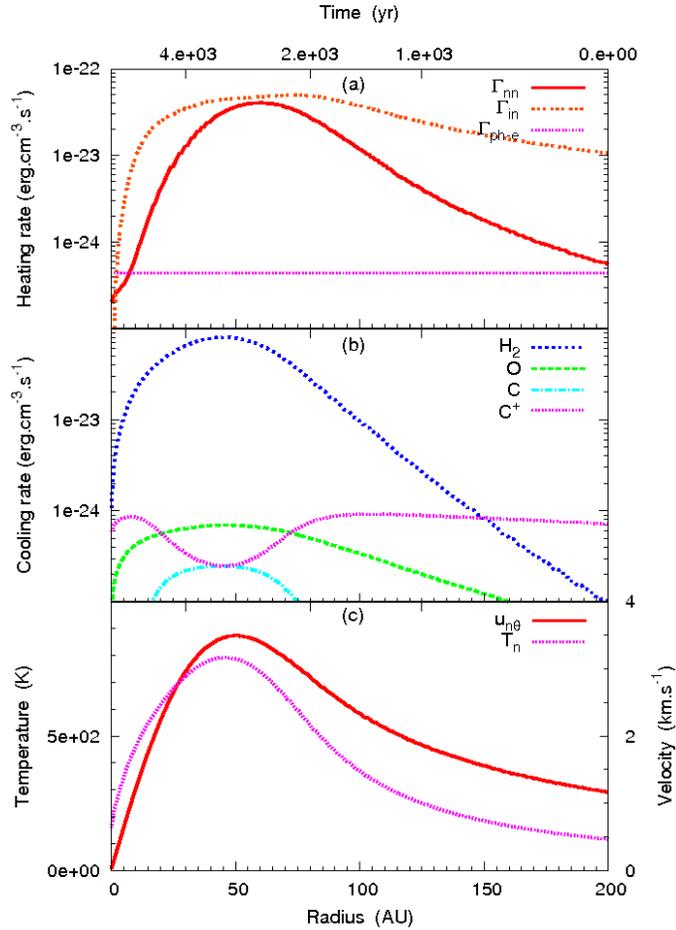}
\caption{Vortex physical properties as functions of the radius
  (bottom axis) or time (top axis, arbitrary origin) for the reference
  model: $a = 3 \times 10^{-11}$ s$^{-1}$, $n_{\rm H} = 30$ cm$^{-3}$ and $A_V
  = 0.4$ mag. Panel (a): The heating terms
  $\Gamma_{nn}$ due to the viscous dissipation, $\Gamma_{in}$ due to the
  ion-neutral friction and $\Gamma_{ph-e}$ due to the photoelectric
  effect. Panel (b): The dominant cooling terms due to the radiative
  desexcitation of H$_2$ (pure rotational lines), C, O and C$^{+}$ 
  (fine-structure lines). Panel (c):
  The temperature and orthoradial velocity of the neutrals.}
\label{FigDyn}
\end{center}
\end{figure}

Fig. \ref{FigDyn} displays the main properties of a reference
model where $a = 3 \times 10^{-11}$ s$^{-1}$, 
$n_{\rm H} = 30$ cm$^{-3}$ and $A_V = 0.4$ mag. 
The vortex has an equilibrium radius $r_0 = 38$ AU
and generates an average 
turbulent heating rate, $\overline{\Gamma}_{turb}= 3.4 \times 10^{-23}$ erg
cm$^{-3}$ s$^{-1}$ defined as:
\begin{equation}
 \overline {\Gamma}_{turb} = 2/(Kr_0)^2  \int_0^{Kr_0}
[ \Gamma_{nn}(r) + \Gamma_{in}(r)] r \,dr.
\end{equation}
The integration domain 
extends to the radius $Kr_0$ where the turbulent heating
has no significant influence on the gas temperature and chemistry ($K\sim 5$).

In the model presented here, the heating rate is dominated everywhere by
ion-neutral friction (Fig. \ref{FigDyn}a). 
For higher values of the turbulent rate of strain $a$, $r_0$ decreases (see
Sect. \ref{LosVarPar}) and  
the importance of the viscous dissipation increases because 
$u_{n\theta}$ is fixed by the ambient turbulence.

The gas in the vortex
 never reaches thermal balance and the thermal inertia is visible
by comparing Figs. \ref{FigDyn}a and \ref{FigDyn}c where the peak temperature
of the fluid particle is reached a few hundred years after the peak of the
heating rate. Emission in the 
pure rotational lines of H$_2$ is by far the dominant coolant in the layers 
where $T_n \gtrsim 200$K (Fig. \ref{FigDyn}c) while  
the cooling rate due to the ionized carbon C$^{+}$ decreases in
the warmest layers. This is due to the chemical evolution (see 
Sect. \ref{ChemVA}).

Last, some chemical clues are provided in Fig. \ref{FigDyn}c. 
The neutral-neutral
reactions only depend on the temperature $T_n$ while the ion-neutral reactions
depend on the ion-neutral drift. In Eq. (\ref{EqTeff}), 
the second term of the right hand side reaches 1000K as $u_D \sim 3$
km s$^{-1}$, an effective temperature higher than 
the peak kinetic temperature in the vortex.  A comparison
of the shapes of the orthoradial velocity $u_{n\theta}$ and the neutral
temperature $T_n$ shows that the endo-energetic 
ion-neutral chemistry is activated earlier in the fluid particle evolution
than the neutral-neutral chemistry. For each type of vortex, 
the relative importance of those two chemical regimes is different.

\subsection{Chemical evolution of the gas} \label{ChemVA}

The chemical evolution of the gas during the vortex active
phase is similar to that reported in J98, although the chemical network 
is updated and includes nitrogen- and sulfur-bearing molecules. 
The outline of this network is given in Appendix \ref{Network} where 
we display the main production and destruction routes 
of the molecules of interest 
(1) in the ambient diffuse medium ($n_{\rm H} = 30$ cm$^{-3}$
and $A_V = 0.4$ mag) and (2) in the vortex for the reference model 
at a radius $r = r_0$.

The most important reaction route opened by the dissipative
structure is the endothermic hydrogenation of C$^{+}$:
\begin{equation} \label{EqCH+}
  \qquad {\rm C}^{+} + {\rm H}_2 \rightarrow {\rm CH}^{+} + {\rm H} \qquad 
-\Delta E/k = 4640 K.
\end{equation}

Besides the direct production of CH$^{+}$, this reaction is responsible
for most of the chemical richness of the vortex as shown in
Fig. \ref{ChemNetwork2} (Appendix \ref{Network}): it enhances the production of
CH$_3^{+}$ via the successive hydrogenation by \HH\ of CH$^{+}$
and CH$_2^{+}$. CH$_3^{+}$, in turn, enhances the production of
molecules including  CH and C via the dissociative recombination with
electrons, HCO$^{+}$ and CO via 
\begin{equation} \label{EqCH3+2}
  \qquad {\rm CH}_3^{+}  + {\rm O} \rightarrow {\rm HCO}^{+} + {\rm H}_2
\end{equation}
and CN, HCN and HNC via
\begin{equation} \label{EqCH3+3}
  \qquad {\rm CH}_3^{+} + {\rm N} \rightarrow {\rm HCN}^{+} + {\rm H}_2.
\end{equation}
The production of CH$_3^{+}$ is also at the origin of C$_2$H and CS since
these molecules are both products of CH (through the reactions CH
+ C$^{+}$ $\rightarrow$ C$_2^{+}$ + H and CH + S$^{+}$ $\rightarrow$ CS$^{+}$
+ H respectively).

The second main reaction, absent in the ambient medium, which
plays an important role in the chemical evolution of the vortex is:
\begin{equation} \label{EqOH}
  \qquad {\rm O} + {\rm H}_2 \rightarrow {\rm OH} + {\rm H}    \qquad - \Delta E/k = 2980 K.
\end{equation}
Besides the direct production of OH it triggers the production of 
H$_2$O via 
\begin{equation} \label{EqH2O}
  \qquad {\rm OH} + {\rm H}_2    \rightarrow {\rm H_2O} + {\rm H} 
\qquad - \Delta E/k = 1490 K
\end{equation}
and O$_2$ via
\begin{equation} \label{EqO2}
  \qquad {\rm OH} + {\rm O}    \rightarrow {\rm O}_2 + {\rm H}.
\end{equation}

\begin{figure}[!h]
\begin{center}
\includegraphics[width=9cm,angle=0]{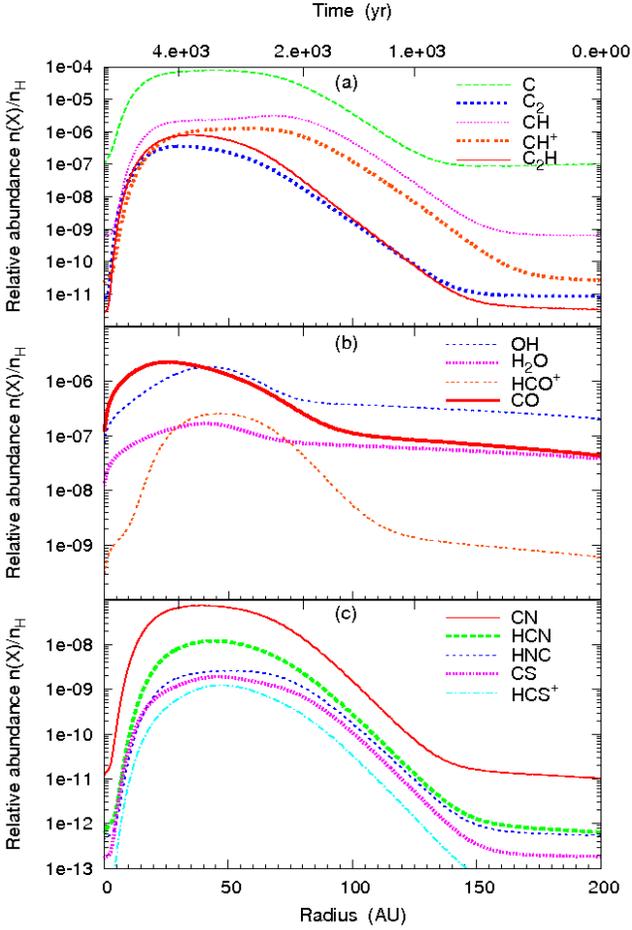}
\caption{Fractional abundances relative to $n_{\rm
  H}=n(\textrm{H})+2n(\textrm{H}_2)$  of selected species as functions of the
  radius (bottom axis) or time (top axis, arbitrary origin) for the
  reference model: $a = 3 \times 10^{-11}$ s$^{-1}$, $n_{\rm H} = 30$
  cm$^{-3}$ and $A_V = 0.4$ mag.}
\label{FigAbrel}
\end{center}
\end{figure}

Fig.\ref{FigAbrel} displays the evolution of several relative abundances
in the magnetized vortex. The impact of the turbulent heating
 is such that most species abundances rise from 2 to 5 orders of
magnitude within the structure. 
The formation of vortices in the turbulent gas flow therefore has specific
chemical signatures that we expect to observe in the diffuse medium. 
One remarkable example is HCO$^{+}$. Since this molecule is a direct
product of CH$_3^{+}$ (in the vortex) it becomes related to almost all the
species whose abundance is enhanced in the vortex. 

Time (and space) stratification is also visible in Fig.\ref{FigAbrel}: the 
chemical enrichments occur successively, because 
the ion-neutral chemistry is
triggered earlier than the neutral-neutral chemistry (see previous
section) and because the chemical inertia of each species is different.

\section{The relaxation phase} \label{VortexRelax}

\subsection{Numerical modelling}

Once the burst of turbulent dissipation is over, some chemical
signatures imprinted in the gas persist for several thousand years as
shown by  Falgarone et al. (2006).
To compute the chemical and thermal relaxation of the gas,  
 we assume that, once the vortex has vanished, the gas is 
dynamically frozen: $\mathbf{u} = 0$. The previous
Lagrangian approach is switched to Eulerian, and
we compute the time-dependent evolution of each cell 
in the vortex. 
The initial conditions of the relaxation are the conditions of
the active stage at every position.
The thermal equations (\ref{EqUn}) and (\ref{EqUi}) are still
valid with $\Gamma_{nn} = 0$ and $\Gamma_{in} = 0$. 

\begin{figure}[!h]
\begin{center}
\includegraphics[width=9cm,angle=0]{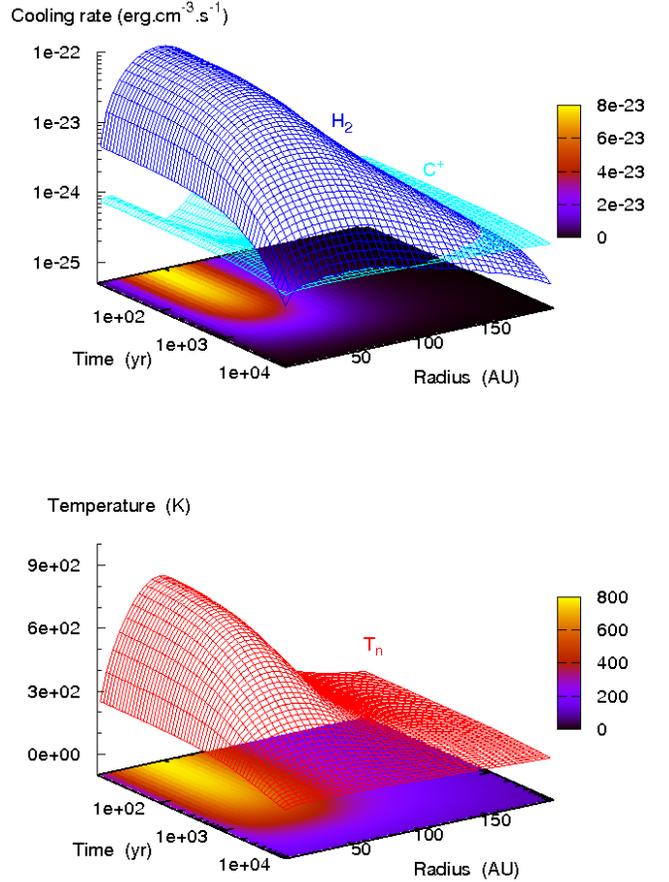}
\caption{Vortex thermal features during the relaxation phase, as functions of
  the radius and time (after the vortex blow-up) for the reference
  model: $a = 3 \times 10^{-11}$ s$^{-1}$, $n_{\rm H} = 30$ cm$^{-3}$ and $A_V = 0.4$
  mag. An isochoric relaxation is assumed.
  Top panel: the main cooling terms, \ie\ the radiative desexcitation of H$_2$
  and C$^{+}$. Bottom panel: the temperature of the neutrals.}
\label{FigCool3D}
\end{center}
\end{figure}

While the numerical code has been conceived to treat 
isobaric or isochoric relaxation, all the results presented in this paper
were obtained assuming isochoric relaxation, because it allows us to better
disentangle what is due to the chemistry itself from what is due to the gas 
density. In particular, it shows more clearly  the role of the 
relaxation timescales of the molecules, \ie\ only driven by the
chemical network and the thermal evolution, independently of the
gas density.

\subsection{Thermal evolution of the gas} \label{ThermVR}

Fig. \ref{FigCool3D} displays the main cooling rates (top panel) and
gas temperature (bottom panel) as functions of time and position
in the vortex (after the vortex blow-up).  It shows that as in the
active phase, the cooling rate during the relaxation phase is
dominated by the emission in the pure rotational lines of H$_2$. In
the model presented here ($n_{\rm H}=30$ cm$^{-3}$), other cooling agents
(mainly C$^{+}$) become dominant at $ t \sim 10^{4}$ yr.

\subsection{Chemical evolution of the gas} \label{ChemVR}

\begin{figure}[!h]
\begin{center}
\includegraphics[width=9cm,angle=0]{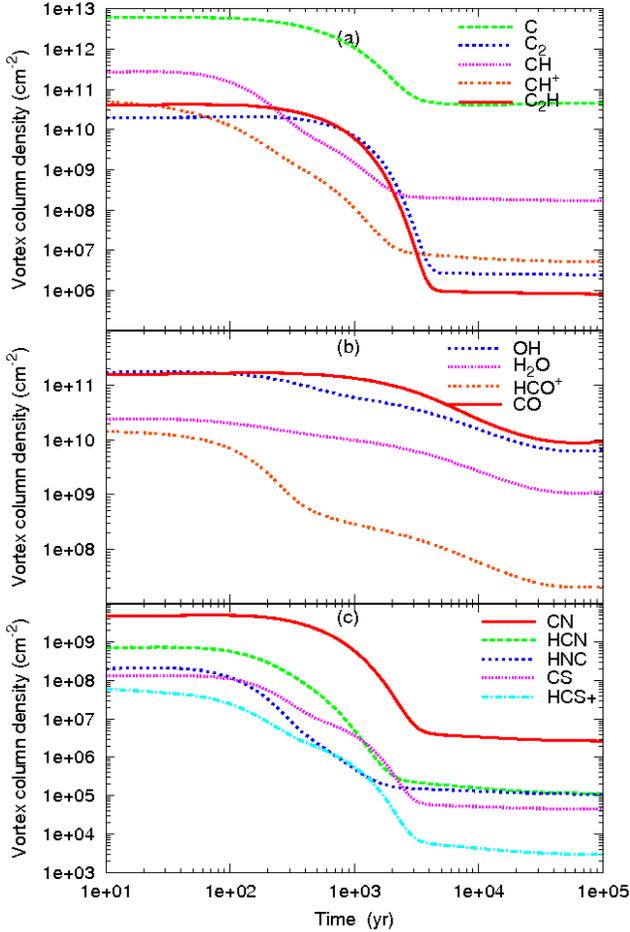}
\caption{Column densities of selected species, integrated across 
 the reference vortex, as functions of time during an isochoric relaxation
 phase.}
\label{FigRelax}
\end{center}
\end{figure}

During relaxation, the cooling of the gas causes 
all the endo-energetic reactions triggered during the dissipative burst
to slow down and stop, one after the other. The gas loses its
chemical enrichment at a speed that depends on the molecular species.
This is illustrated in Fig. \ref{FigRelax} that displays 
the time-dependent evolution of the  column density $N_{VR}(X,t)$ of selected
species $X$ integrated across the vortex. 

For most species (CO, C$_2$, OH, H$_2$O, C$_2$H...) the signature of
the turbulent dissipation persists over more than $10^{3}$ yr. The
characteristic timescales (e-folding times) differ between
species by more than a factor of 30 ($\approx 2 \times 10^{2}$ yr for
CH$^{+}$ and $7 \times 10^{3}$ yr for CO).  The existence of the
relaxation phases modifies the correlations between molecular
abundances. One example is provided by HCN and C$_2$H
that follow a similar enhancement in the active phase (see
Fig. \ref{FigAbrel}) and have markedly different behaviours in the
relaxation phase. 

The richest phases are not necessarily those contributing the most to
the observable column densities because they have short
lifetimes. In the next section, we detail  how we take time-variability
into account in our modelling of a random line of sight across the
diffuse medium.

\section{Modelling of a line of sight} \label{LineOfSight}

In turbulent flows, the spatial distribution of the
regions of highest dissipation rate (extrema of velocity shear, extrema of 
negative velocity divergence \ie\ shocks) is far from space-filling, one of
the aspects of the intermittent nature of turbulence. The filling
factor of these extrema has been computed in numerical simulations of
mildly compressible (Pety \& Falgarone 2000) or supersonic MHD
turbulence (Pavlovski et al. 2006, Pan \& Padoan 2008). More
recently, Moisy \& Jimenez (2004) have shown that the regions of
intense vorticity tend to form filaments, while regions of most
intense dissipation rather form sheets or ribbons, all of them being
organized in clusters, probing the organization of small-scale 
intermittent structures.

For our purpose, we assume that any line of sight intercepts a
number of vortices, either active or in their relaxation phase. The
total number of vortices per line of sight is constrained by the average
transfer rate of turbulent energy per unit volume available in the cascade.

\subsection{Method} \label{LosMethod}

Any line of sight samples three kinds of diffuse
gas: (1) the ambient medium in which the chemistry is 
computed as steady state UV-dominated chemistry, (2) a
number $\mathcal{N}_{VA}$ of active vortices
and (3) a number of relaxation phases related to 
$\mathcal{N}_{VA}$.
We also consider that the line of sight is homogeneous and characterized
by its density $n_{\rm H}$ and uniform shielding
$A_V$ from the ISRF. Such an 
approximation is useful to
test the importance of each parameter ($a$, $n_{\rm H}$ and $A_V$) on the
final chemical state of the gas.
Last, we assume that all the vortices in the line of sight are identical:
they all have the same turbulent rate of strain $a$. In the following section
we show why a more realistic description, with a distribution of
rate of strain values, would not provide very different results, namely
because, as it will be seen, the results depend weakly on $a$.

The contribution of one active phase of duration $\tau_V$ 
to the total column density of a species $X$ is $N_{VA}(X)$. The contribution
of the relaxation phase of that species is computed by assuming that the
longer its relaxation timescale, the greater the contribution of the relaxation
phase in the observed column density so that the chemical composition of a
line of sight is entirely determined by the number of active vortices
$\mathcal{N}_{VA}$ and their lifetime $\tau_V$. The resulting column density
of a species $X$ is:
\begin{equation} \label{EqColDens}
N(X) = \mathcal{N}_{VA}(X) \left[ N_{VA}(X) + 1/\tau_V 
  \int_0^{\infty} N_{VR}(X,t)dt \right] + N_M(X)
\end{equation}
where $N_M(X)$ is the contribution of the ambient medium.

\subsection{Constraints provided by the energy available in the turbulent
  cascade} \label{LosConstraints}

\begin{figure*}[!ht]
\begin{center}
\includegraphics[width=12.5cm,angle=0]{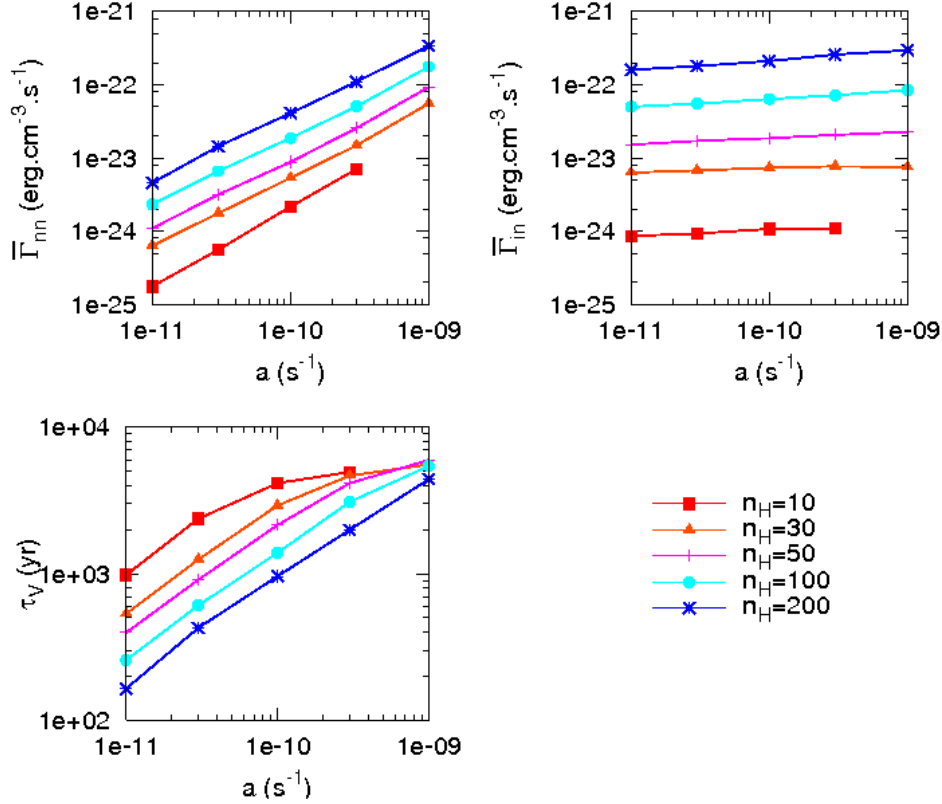}
\caption{Main physical properties of the TDR models as functions of the
  turbulent rate of strain $a$ for several values of the density $n_{\rm H}$.
  Top panels: the vortex heating terms $\overline{\Gamma}_{nn}$ due to the
  viscous dissipation (left) and $\overline{\Gamma}_{in}$ due to the
  ion-neutral friction (right).
  Bottom panel: the vortex lifetime $\tau_V$.}
\label{FigVarParam}
\end{center}
\end{figure*}

\begin{figure*}[!ht]
\begin{center}
\includegraphics[width=12.5cm,angle=-90]{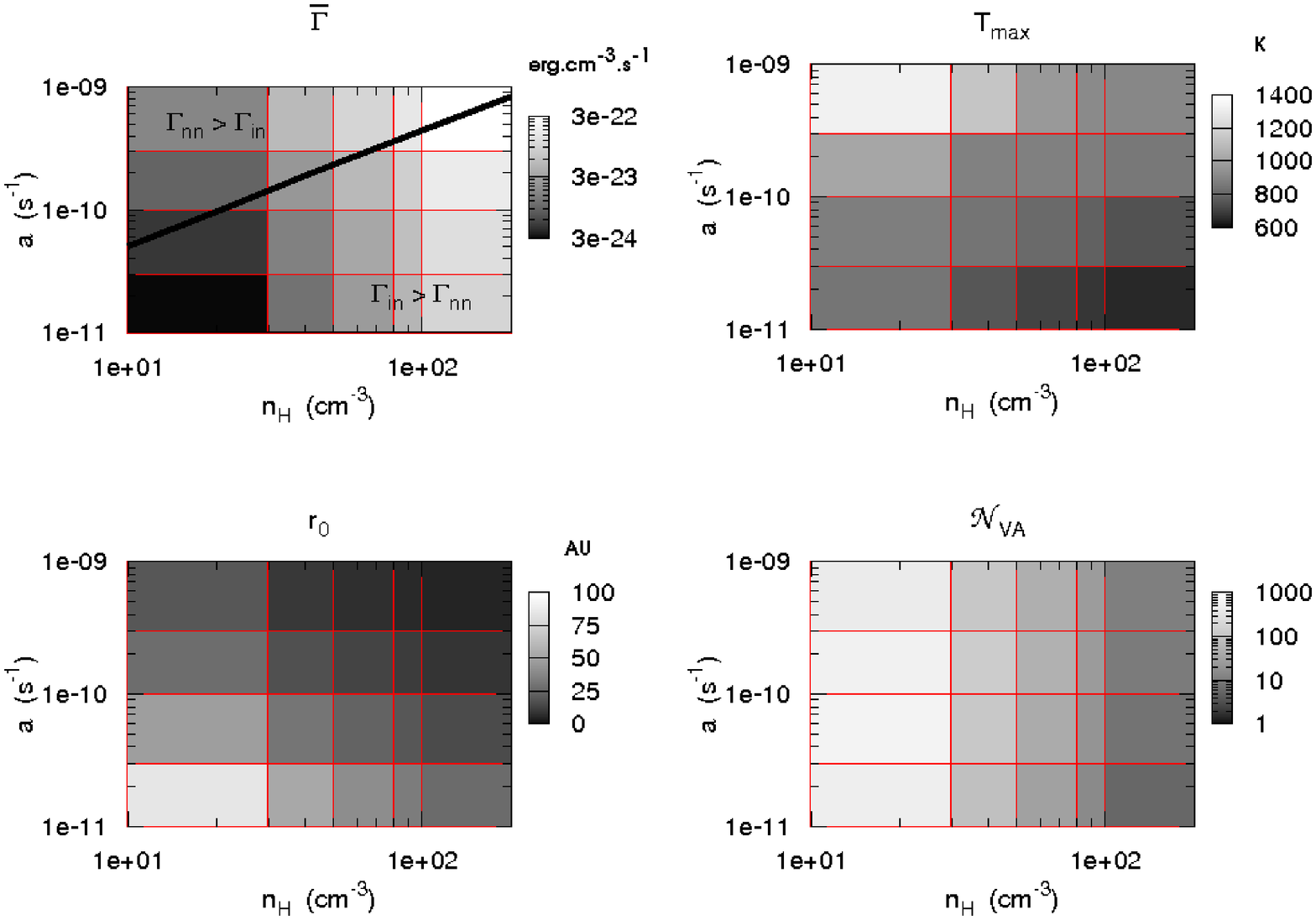}
\caption{Main physical properties of the TDR models as functions of the
  turbulent rate of strain $a$ and of the density $n_{\rm H}$ of the gas.
  Top panels: the vortex turbulent heating rate $\overline{\Gamma}_{turb}$
  (left) and the vortex maximum temperature $T_{max}$ (right).
  Bottom panels: the vortex radius $r_0$ (left) and the number of active
  vortices $\mathcal{N}_{VA}$ along a line of sight sampling one magnitude of
  gas (right).}
\label{FigVarParam2}
\end{center}
\end{figure*}

\subsubsection{The number of active vortices}

The number of active vortices $\mathcal{N}_{VA}$ in
a line of sight and their lifetime $\tau_V$, are 
constrained by the turbulent energy available in the cascade and its 
transfer rate.
In turbulent flows, the transfer rate of kinetic energy 
at scale $l$ of characteristic velocity $u_l$ is, 
on average: 
\begin{equation} \label{EqTrrate}
\varepsilon_l = \frac{1}{2} \rho \frac{u_l^{3}}{l}.
\end{equation} 
The Kolmogorov scaling of turbulence, valid for incompressible
turbulence, postulates that this transfer rate is scale-independent.
In the highly compressible interstellar medium, one would expect this
quantity to differ from one scale to another.  Measurements of the
internal velocity dispersion of clouds of size $l$ combined with
their density provide an estimate of the turbulent energy transfer
rate at this scale. A compilation of CO(1-0) line observations of
interstellar structures of size ranging between $10^{-2}$ and
$10^{3}$ pc shows that $\varepsilon_l$ is remarkably independent of
the scale in the Galaxy and that there is a large scatter (by a
factor of 100) about a well defined average value
${\overline \varepsilon}_{obs} \sim 2 \times 10^{-25}$ \eccs\ (Falgarone,
1998; Hily-Blant et al. 2008). 
A similar value holds for turbulence in the HI CNM and in
non-star-forming dense cores (Falgarone 1999). 
The uniformity of this value
across the local ISM suggests that the turbulent cascade encompasses
the different regions of the cold medium, and that the transfer is
driven by turbulence at the same rate, in all media, whatever the
gas density. 

We adopt the above value of ${\overline \varepsilon}_{obs}$ in our model as 
representative of the turbulent energy transfer rate through
scales. We thus impose that, at any time, the dissipation rate 
in all the active 
vortices in a line of sight is equal to the average energy transfer rate 
in the turbulent cascade, so that:
\begin{equation} \label{EqNVA}
{\overline \varepsilon}_{obs} = 
\mathcal{N}_{VA} \overline{\Gamma}_{turb} 2 K r_0/L
\end{equation}
where $L$ is the depth of the line of sight, inferred from $L = N_{\rm H}/n_{\rm H}$.
This fixes the number of active vortices in a given line of sight\footnote{
  $\mathcal{N}_{VA}$ also depends on the angle of inclination of the
  vortices along the line of sight. Because the dynamics and the chemistry
  in a vortex do not depend on the axial coordinates $z$ and because the model
  is axisymmetric, this angle has no influence on the final
  results. $\mathcal{N}_{VA}$ is thus defined for vortices perpendicular
  to the line of sight.}.
This number and many of the results are therefore  
proportional to ${\overline \varepsilon}_{obs}$.

\subsubsection{The velocity dispersion of the ambient turbulence}

The energy transfer rate depends on the density, velocity
dispersion and timescale. We thus need to know the amount of turbulent
energy available in the CNM component of the diffuse medium, 
or the rms turbulent velocity in the CNM. It is this quantity 
that sets the angular velocity of the vortex (see Section 2.2). This
quantity is difficult to determine on observational grounds, because
of the mixture of WNM and CNM in the HI emission spectra and because of 
the lack
of spatial information for the absorption spectra (dominated by the
CNM).  We adopted a rms velocity dispersion $\sigma_{turb} = 3.5$ \kms\
for the CNM turbulence derived from the HI maps of a high latitude
cirrus in the \object{Polaris Flare} (Joncas et al. 1992; Miville-Desch\^enes
et al. 2003) in which
HI emission is well correlated with the 100$\mu$m emission of dust,
probing column densities of gas representative of the CNM. This value
is consistent with those quoted in Crovisier (1981) for the 
CNM. It is comparable
to the geometric mean of the two smallest values, $FWHM=4.9$ and 12.0
\kms\ inferred by Haud \& Kalberla (2007) from the Gaussian
decomposition of the HI profiles of the Leiden/Argentine/Bonn survey
of galactic HI (Hartmann \& Burton 1997). Last, this rms velocity
dispersion is consistent with the approximate equipartition
between magnetic and turbulent energy inferred by Heiles \& Troland
(2005) from the median value of the magnetic field estimated in the
diffuse medium $B = 6$ $\mu$G. 

\subsubsection{The lifetime of the active phase} \label{lifetime}

The lifetime of an active vortex, $\tau_V$, is controlled by the
large-scale motions of the ambient turbulence that feed energy into the 
small-scale structures (Moffatt, Kida \& Ohkitani 1994). It may exceed
the period of the vortex as suggested by
a variety of experiments in incompressible turbulence (\eg\ Douady, Couder,
Brachet 1991). This lifetime is uncertain, though, and  its value in our model
is constrained by energetic considerations.

We assume that, for simplicity, all the vortices explored in the models
dissipate the same total energy over their lifetime $\tau_V$

\begin{equation} \label{EqTotEnergy}
E = \pi (Kr_0)^{2} \, \overline{\Gamma}_{turb} \, L_V \, \tau_V
\end{equation}

This constraint fixes the lifetime $\tau_V$ of the vortex \ie\ the time
during which turbulent dissipation is active. In order to stay in the vortex
framework we impose that $\tau_V$ is larger than the vortex period $P$, which
sets a lower limit to $E$. The influence of this parameter on the results is
discussed in Appendix \ref{ImpactTauv}.

\subsection{The role of the rate of strain and density under the energetic 
constraints} \label{LosVarPar}

Once ${\overline \varepsilon}_{obs}$, $\sigma_{turb}$ and $E$ are
given, a line of sight is therefore defined by only three independent
parameters: the turbulent rate of strain $a$, the gas density $n_{\rm H}$
and the shielding $A_V$.  

To help the reader understand the chemical results presented in the
next section, we discuss the roles of the
rate of strain and gas density because the above constraints on the 
energy dissipation rate actually couple  
$a$ and $n_{\rm H}$ that should be independent parameters. This is so
because the two energy constraints (transfer rate and energy)
involve the equilibrium radius $r_0$ that primarily 
depends on $a$ but also on $n_{\rm H}$, via
the  density dependence of the
kinematic viscosity (see Section 2.1)
These trends are illustrated in 
Figs. \ref{FigVarParam} and \ref{FigVarParam2} 
that also display the dependence of several 
key quantities on $a$ and $n_{\rm H}$.

Fig. \ref{FigVarParam} shows that, as expected, $\Gamma_{nn} \sim a
 n_{\rm H}$ is almost proportional to $a$ because the higher the
 rate of strain, the smaller the equilibrium radius and the larger the
 velocity shear.
 $\Gamma_{in}$ is almost independent of $a$ and increases with density
 as $\Gamma_{in} \propto u_D n_{\rm H}^{2}$ so that, depending on $a$
 and $n_{\rm H}$, two regimes exist: one at low density and high
 rate of strain where the turbulent heating is dominated by viscous
 dissipation, and the other (small $a$, high density) where this
 heating is dominated by the ion-neutral friction.
The rate of strain $a$ therefore plays an
 important role in the nature of the warm chemistry triggered in the
 vortex. 
This figure also displays the run of $\tau_V$ with 
the rate of strain for different densities, as a result of the constraint
on the total energy $E$ dissipated in each vortex.

Fig.\ref{FigVarParam2} shows that $r_0$ is small in the former regime
 and reaches values of the order of 100 AU in the regime where
 ion-neutral friction dominates.  
The peak gas temperature reached in the vortex, $T_{max}$, is also
 shown: it is higher in the regime where viscous dissipation dominates
 because the orthoradial velocity is fixed and thin vortices induce
 large velocity shears, thus large viscous heating. However, the
 thermal inertia of the gas prevents it from reaching much higher
 temperatures, because the most efficient vortices (large $a$) 
are short-lived.  The range of $a$ and $n_{\rm H}$
 explored in our study covers these two regimes and we quantify the
 chemical effects of turbulent dissipation as it changes from
 dominated by ion-neutral friction to dominated by viscous dissipation.

Last, we find that the number of active vortices $\mathcal{N}_{VA}$ (computed
for a line of sight sampling one magnitude of gas) is
roughly independent of $a$ and decreases almost as $n_{\rm H}^{-2}$ as the
density increases. This is because $\overline{\varepsilon}_{obs}$ is fixed
(see previous section), and because of the combined dependences of $r_0$,
$\overline{\Gamma}_{turb}$ and $L$ on $a$ and $n_{\rm H}$ (see Eq. \ref{EqNVA}).
We note that $\mathcal{N}_{VA}$ reaches 
large values at low density (up to several hundreds along lines of sight 
of several tens of parsecs). However the filling factor of the vortices
\begin{equation}
f_v = \mathcal{N}_{VA} \frac{2Kr_0}{L}
\end{equation}
never exceeds $f_v = 4 \times 10^{-2}$, its lowest values being $f_v \sim 
 10^{-4}$ at high densities and rates of strain.

\subsection{The respective contribution of each chemical regime} \label{contributions}

\begin{table*}
\begin{center}
\caption{Physical and chemical characteristics of two TDR-models defined by
  their parameters $n_{\rm H}$, $A_V$ and $a$. $f_M(X)$, $f_{VA}(X)$
  and $f_{VR}(X)$ are the contributions of the ambient
  medium, the active and the relaxation phases respectively, to the
  column density $N(X)$ of the species $X$. All models are computed for $N_{\rm
  H} = 1.8 \times 10^{21} \rm{cm}^{-2}$. The relative abundances are given in
  the last columns. Numbers in parenthesis are powers of 10.}
\begin{tabular}{l | c c c c c | c c c c c }
\hline
\multicolumn{11}{c}{Parameters} \\
\hline
$n_{\rm H}$ & \multicolumn{5}{c|}{30 cm$^{-3}$}    & \multicolumn{5}{c}{100 cm$^{-3}$} \\
$A_V$ & \multicolumn{5}{c|}{0.1 mag}         & \multicolumn{5}{c}{0.1 mag} \\
$a$   & \multicolumn{5}{c|}{3 (-11) s$^{-1}$}& \multicolumn{5}{c}{3 (-11) s$^{-1}$} \\
\hline
\multicolumn{11}{c}{Physical properties} \\
\hline
$\overline{\Gamma}_{turb}$ &
\multicolumn{5}{c|}{9.3 (-24) erg.cm$^{-3}$.s$^{-1}$} & 
\multicolumn{5}{c }{6.9 (-23) erg.cm$^{-3}$.s$^{-1}$} \\
$T_{amb}$ & \multicolumn{5}{c|}{ 114 K} & \multicolumn{5}{c}{ 58 K} \\
$T_{max}$ & \multicolumn{5}{c|}{ 1000 K} & \multicolumn{5}{c}{807 K} \\
$\tau_V$  & \multicolumn{5}{c|}{1070 yr} & \multicolumn{5}{c}{506 yr} \\
$r_0$     & \multicolumn{5}{c|}{ 40 AU} & \multicolumn{5}{c}{ 21 AU} \\
$\mathcal{N}_{VA}$ & \multicolumn{5}{c|}{ 215} & \multicolumn{5}{c}{ 17} \\
$f_{v}$            & \multicolumn{5}{c|}{ 2.2 (-2)} & \multicolumn{5}{c}{ 2.9 (-3)} \\
\hline
\multicolumn{11}{c}{Chemical properties} \\
\hline
&
$f_M(X)$ &
$f_{VA}(X)$ &
$f_{VR}(X)$ &
$N(X)$ &
$N(X)/N_{\rm H}$ &
$f_M(X)$ &
$f_{VA}(X)$ &
$f_{VR}(X)$ &
$N(X)$ &
$N(X)/N_{\rm H}$ \\
Species & \% & \% & \% & cm$^{-2}$ & & \% & \% & \% & cm$^{-2}$ & \\
\hline
H            & 100 & 0$^{a}$ & 0$^{a}$ & 7.0 (20) &  3.9 (-01) & 100 & 0$^{a}$ & 0$^{a}$ & 2.8 (20) & 1.6 (-02)\\
H$_2$        & 100 & 0$^{a}$ & 0$^{a}$ & 5.5 (20) &  3.1 (-01) & 100 & 0$^{a}$ & 0$^{a}$ & 7.6 (20) & 4.2 (-01)\\
H$_3^{+}$    &  81 &  5 & 14 & 2.9 (13) &  1.6 (-08) &  93 &  2 &  5 & 1.2 (13) & 6.7 (-09)\\

C            &  19 & 65 & 35 & 4.4 (14) &  2.4 (-07) &  68 & 23 &  9 & 5.6 (14) & 3.1 (-07)\\
CH           &   5 & 86 &  9 & 8.1 (12) &  4.5 (-09) &  22 & 69 &  9 & 7.9 (12) & 4.4 (-09)\\
CH$^{+}$     & 0.2 & 96 &  4 & 2.2 (13) &  1.2 (-08) &   2 & 95 &  3 & 1.3 (12) & 7.2 (-10)\\
C$_2$        & 0.5 & 72 & 28 & 4.0 (11) &  2.2 (-10) &   3 & 46 & 51 & 1.4 (12) & 7.8 (-10)\\
C$_2$H       & 0.1 & 76 & 24 & 6.3 (11) &  3.5 (-10) & 0.4 & 61 & 39 & 3.5 (12) & 1.9 (-09)\\

OH           &  52 & 17 & 31 & 8.7 (13) &  4.8 (-08) &  63 & 18 & 19 & 2.3 (13) & 1.3 (-08)\\
H$_2$O       &  57 & 11 & 32 & 1.3 (13) &  7.2 (-09) &  68 & 14 & 18 & 3.6 (12) & 2.0 (-09)\\
H$_3$O$^{+}$ &  35 & 23 & 42 & 1.3 (13) &  7.2 (-09) &  32 & 40 & 28 & 1.1 (12) & 6.1 (-10)\\
CO           &  57 & 16 & 27 & 2.6 (13) &  1.4 (-08) &  67 & 12 & 21 & 3.0 (13) & 1.7 (-08)\\
HCO$^{+}$    &  14 & 69 & 17 & 8.4 (11) &  4.7 (-10) &  11 & 77 & 12 & 3.7 (11) & 2.1 (-10)\\
O$_2$        &  67 & 11 & 22 & 3.8 (10) &  2.1 (-11) &  80 & 10 & 10 & 4.1 (10) & 2.3 (-11)\\

SH$^{+}$     & 0.1 & 98 &  2 & 6.2 (11) &  3.4 (-10) & 0.2 & 98 &  2 & 2.1 (11) & 1.2 (-10)\\
CS           &   3 & 85 & 12 & 1.8 (09) &  1.0 (-12) &  11 & 69 & 20 & 7.0 (09) & 3.9 (-12)\\
HCS$^{+}$    & 0.6 & 93 &  7 & 1.1 (09) &  6.1 (-13) &   1 & 89 & 10 & 1.9 (09) & 1.1 (-12)\\

CN           &   1 & 79 & 20 & 2.3 (11) &  1.3 (-10) &   8 & 63 & 29 & 1.6 (11) & 8.9 (-11)\\
HCN          & 0.5 & 88 & 12 & 3.3 (10) &  1.8 (-11) &   3 & 81 & 16 & 2.6 (10) & 1.4 (-11)\\
HNC          &   2 & 89 &  9 & 5.9 (09) &  3.3 (-12) &  17 & 74 &  9 & 4.0 (09) & 2.2 (-13)\\

\hline
\end{tabular}
\begin{list}{}{}
\item[$a$] For the reason given in Sect. \ref{contributions}
  $f_{VA}({\rm H})= f_{VR}({\rm H}) = f_{VA}({\rm H_2}) = f_{VR}({\rm H_2}) = 0$,
  because the densities of H and H$_2$ are not modified in the vortex.
\end{list}
\label{TabTriangles1}
\end{center}
\end{table*}

\begin{table*}
\begin{center}
\caption{ Same as Table \ref{TabTriangles1}.}
\begin{tabular}{l | c c c c c | c c c c c }
\hline
\multicolumn{11}{c}{Parameters} \\
\hline
$n_{\rm H}$ & \multicolumn{5}{c|}{30 cm$^{-3}$}    & \multicolumn{5}{c}{100 cm$^{-3}$} \\
$A_V$ & \multicolumn{5}{c|}{0.4 mag}         & \multicolumn{5}{c}{0.4 mag} \\
$a$   & \multicolumn{5}{c|}{3 (-11) s$^{-1}$}& \multicolumn{5}{c}{3 (-11) s$^{-1}$} \\
\hline
\multicolumn{11}{c}{Physical properties} \\
\hline
$\overline{\Gamma}_{turb}$ &
\multicolumn{5}{c|}{8.5 (-24) erg.cm$^{-3}$.s$^{-1}$} & 
\multicolumn{5}{c }{6.2 (-23) erg.cm$^{-3}$.s$^{-1}$} \\
$T_{amb}$ & \multicolumn{5}{c|}{ 79 K} & \multicolumn{5}{c}{ 42 K} \\
$T_{max}$ & \multicolumn{5}{c|}{ 830 K} & \multicolumn{5}{c}{710 K} \\
$\tau_V$  & \multicolumn{5}{c|}{1250 yr} & \multicolumn{5}{c}{610 yr} \\
$r_0$     & \multicolumn{5}{c|}{ 38 AU} & \multicolumn{5}{c}{ 20 AU} \\
$\mathcal{N}_{VA}$ & \multicolumn{5}{c|}{ 246} & \multicolumn{5}{c}{ 19} \\
$f_{v}$            & \multicolumn{5}{c|}{2.4 (-2) } & \multicolumn{5}{c}{3.2 (-3)} \\
\hline
\multicolumn{11}{c}{Chemical properties} \\
\hline
&
$f_M(X)$ &
$f_{VA}(X)$ &
$f_{VR}(X)$ &
$N(X)$ &
$N(X)/N_{\rm H}$ &
$f_M(X)$ &
$f_{VA}(X)$ &
$f_{VR}(X)$ &
$N(X)$ &
$N(X)/N_{\rm H}$ \\
Species & \% & \% & \% & cm$^{-2}$ & & \% & \% & \% & cm$^{-2}$ & \\
\hline
H            & 100 & 0$^{a}$ & 0$^{a}$ & 2.0 (20) & 1.1 (-01) & 100 & 0$^{a}$ & 0$^{a}$ & 7.9 (19) & 4.4 (-02)\\
H$_2$        & 100 & 0$^{a}$ & 0$^{a}$ & 8.0 (20) & 4.4 (-01) & 100 & 0$^{a}$ & 0$^{a}$ & 8.6 (20) & 4.8 (-01)\\
H$_3^{+}$    &  77 &  6 & 17 & 5.9 (13) & 3.3 (-08) &  92 &  2 &  6 & 1.4 (13) & 7.8 (-09)\\

C            &  19 & 56 & 25 & 1.3 (15) &  7.2 (-07) &  76 & 13 & 11 & 1.4 (15) & 7.8 (-07)\\
CH           &   2 & 87 & 11 & 3.6 (13) &  2.0 (-08) &  20 & 66 & 14 & 1.5 (13) & 8.3 (-09)\\
CH$^{+}$     & 0.2 & 96 &  4 & 1.3 (13) &  7.2 (-09) &   3 & 93 &  4 & 6.0 (11) & 3.3 (-10)\\
C$_2$        & 0.3 & 57 & 43 & 4.2 (12) &  2.3 (-09) &   3 & 30 & 67 & 4.2 (12) & 2.3 (-09)\\
C$_2$H       & 0.1 & 68 & 32 & 7.4 (12) &  4.1 (-09) & 0.4 & 52 & 47 & 1.0 (13) & 5.6 (-09)\\

OH           &  47 & 24 & 29 & 8.4 (13) &  4.7 (-08) &  34 & 31 & 35 & 1.5 (13) & 8.3 (-09)\\
H$_2$O       &  50 & 20 & 30 & 1.3 (13) &  7.2 (-09) &  39 & 27 & 34 & 2.4 (12) & 1.3 (-09)\\
H$_3$O$^{+}$ &  20 & 46 & 34 & 9.9 (12) &  5.5 (-09) &  11 & 59 & 30 & 7.0 (11) & 3.9 (-09)\\
CO           &  45 & 15 & 40 & 1.2 (14) &  6.7 (-08) &  35 & 11 & 54 & 9.1 (13) & 5.1 (-08)\\
HCO$^{+}$    &   6 & 82 & 12 & 2.2 (12) &  1.2 (-09) &   3 & 84 & 13 & 4.9 (11) & 2.7 (-10)\\
O$_2$        &  64 & 16 & 20 & 6.9 (10) &  3.8 (-11) &  52 & 19 & 29 & 4.4 (10) & 2.4 (-11)\\

SH$^{+}$     & 0.02& 97 &  3 & 1.7 (12) &  9.4 (-10) & 0.03& 96 &  4 & 4.9 (11) & 2.7 (-10)\\
CS           &   1 & 82 & 17 & 1.9 (10) &  1.1 (-11) &   9 & 61 & 30 & 3.0 (10) & 1.7 (-11)\\
HCS$^{+}$    & 0.2 & 92 &  8 & 9.2 (09) &  5.1 (-12) &   1 & 86 & 13 & 6.1 (09) & 3.4 (-12)\\

CN           &   2 & 67 & 31 & 8.8 (11) &  4.9 (-10) &  10 & 46 & 44 & 4.7 (11) & 2.6 (-10)\\
HCN          &   1 & 84 & 15 & 1.0 (11) &  5.6 (-11) &   3 & 75 & 2  & 5.7 (10) & 3.2 (-11)\\
HNC          &   2 & 87 & 11 & 2.7 (10) &  1.5 (-11) &  18 & 69 & 13 & 1.2 (10) & 6.7 (-12)\\

\hline
\end{tabular}
\begin{list}{}{}
\item[$a$] Same as Table \ref{TabTriangles1}.
\end{list}
\label{TabTriangles2}
\end{center}
\end{table*}

Three different chemical regimes enter our line of sight modelling, associated
with the active phase, the relaxation phase and the ambient medium. Their
respective contributions to the integrated column density of a species $X$
(see Eq. \ref{EqColDens}),
\begin{equation}
f_M(X) = N_M(X)/N(X)
\end{equation}
\begin{equation}
f_{VA}(X)=\mathcal{N}_{VA}(X) N_{VA}(X) / N(X)
\end{equation}
\begin{equation}
f_{VR}(X)=\mathcal{N}_{VA}(X) \int_0^{\infty} N_{VR}(X,t)dt / N(X)\tau_V
\end{equation}
depend on this species and are given in Tables. \ref{TabTriangles1} \&
\ref{TabTriangles2}.

For numerical reasons the contributions $f_{VA}(X)$ and $f_{VR}(X)$ are
computed as excesses of the density $n(X)$ in the active and relaxation phases
respectively, above its value in the ambient medium. Hence $N_M(X) = f_{M}(X)
\times N(X)$ is the column density of the species $X$ along a line of sight
without vortices of total column density $N_{\rm H} = 1.8 \times 10^{21}$
cm$^{-2}$.

Tables. \ref{TabTriangles1} \& \ref{TabTriangles2} show that all the
regimes have a significant contribution, although most of the selected species
are predominantly formed in the vortices.

\section{Results: Comparison with the observations} \label{Results}

We have computed 150 different models exploring the parameter space as 
follows:\\
- 6 densities ($n_{\rm H}=$ 10, 30, 50, 80, 100, 200 \cc)\\
- 5 rates of strain ($a=$ 1, 3, 10, 30, 100 $\times 10^{-11}$ s$^{-1}$)\\
- 5 UV shieldings ($A_V= 0.2$, 0.4, 0.6, 0.8, 1).\\
The turbulent rate of strain extends over two orders of magnitude, the
largest value corresponding to the condition $r_0 \gg \lambda_{\rm
HH}$ for all densities, the smallest being the limit of validity of
our analytical approach.

In this Section, we present the results of our line of sight models 
(named Turbulent Dissipation Regions models or TDR models) on
the same displays as large sets of observational data. We also compare
these data with results obtained from PDR models (two-side
illuminated slabs of uniform density (Meudon PDR code, Le Petit et
al. 2006)), computed with the same conditions as in the TDR models: $\chi=1$,
$\zeta = 3\times 10^{-17}$ s$^{-1}$ and with the same chemical network.

The amount of gas along the line of sight in the PDR and TDR models is
normalized to $N_{\rm H} = 1.8 \times 10^{21}$
cm$^{-2}$  because, as will be shown later, it corresponds to most of the data
collected in the visible,  UV and radio ranges.

\subsection{Ultraviolet and visible observations towards nearby stars} \label{ObsUV}
An important clue for the understanding of the diffuse ISM chemistry
is provided by the combined observations of CH and CH$^{+}$ (
Gredel et al. 1993; Crane et
al. 1995 ; Gredel 1997; Pan et al. 2004, 2005; Sheffer et al. 2008).
Although CH$^{+}$ is known to be overabundant, the column densities of CH are
in agreement with the predictions of the PDR models. Because the two
species are tightly related by the chemistry (see Figs. \ref{ChemNetwork1} \&
\ref{ChemNetwork2} in Appendix \ref{Network}), the difficulty of any modelling
is then to understand the physics which leads to an enhancement of CH$^{+}$
without changing the amount of CH.  Another important clue lies in  
the correlation, mentioned in the Introduction, between CH$^{+}$ and the pure
rotational $J\geqslant3$ levels of H$_2$.

\begin{figure*}[!ht]
\begin{center}
\includegraphics[width=14.8cm,angle=0]{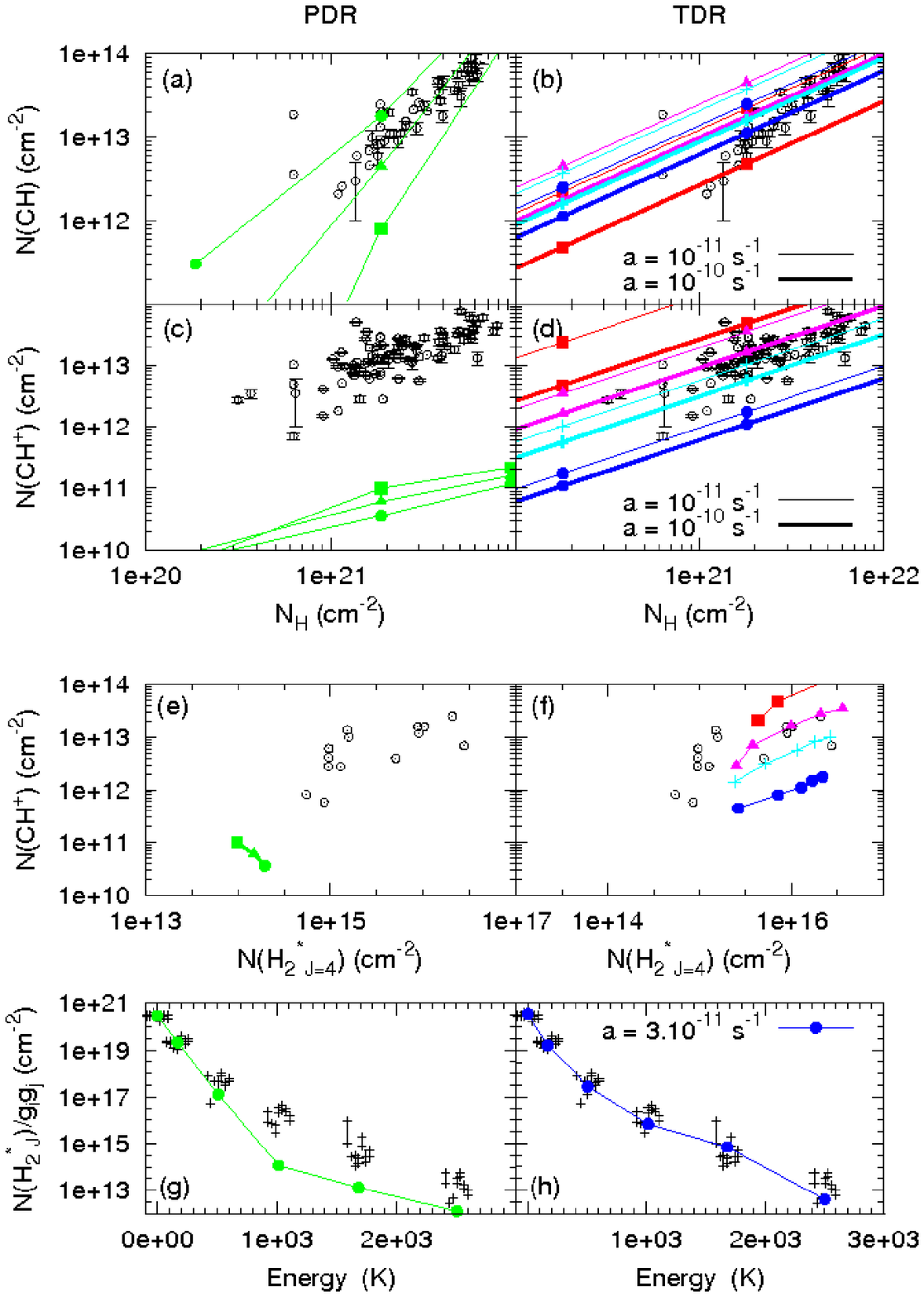}
\caption{Observations compared to PDR models (left panels) and to
  TDR models (right panels). {\bf Data} (open circles) - CH column densities
  are from Crane et al. (1995) and Gredel (1997). CH$^{+}$ column densities
  are from Crane et al. (1995), Gredel (1997) and Weselak et
  al. (2008). H$_{2\,J=4}^{*}$ column densities are from Spitzer et
  al. (1974), Snow (1976, 1977), Frisch (1980), Frisch \& Jura
  (1980) and Lambert \& Danks (1986).
  The excitation diagrams of H$_2$ are from Gry et
  al. (2002) and Lacour et al. (2005). The data on panels (e,f) and (g,h) are
  scaled to $N_{\rm H} = 1.8 \times 10^{21}$ cm$^{-2}$ and $N(\rm{H_2}) =
  5\,\,10^{20} \rm{cm}^{-2}$ respectively, and correspond to lines of sight
  toward stars of type later than B2. For clarity, on panels (g,h), the points
  for a given level are slightly shifted on the abscissa.
  {\bf PDR and TDR models} (filled symbols) - Computed for
  several densities: 10 (squares), 30 (triangles), 50 (crosses) and 100
  cm$^{-3}$ (circles). All the TDR models are computed for $A_V=0.2$.
  The models of panels (e,f) and (g,h) have been scaled as the data. 
  In panel (f) the TDR models are computed
  for $a$ varying along each curve between $10^{-11}$ (top right) and
  $10^{-9}$ s$^{-1}$ (bottom left).}
\label{FigCorrelationUV}
\end{center}
\end{figure*}

The CH$^{+}$, CH and excited H$_2$ observational data are displayed
in Fig. \ref{FigCorrelationUV} with the prediction of several
PDR models (left panels) and TDR models (right panels), computed for
diffuse gas of density between 10 and 200 cm$^{-3}$.
As mentioned above, we restrict our analysis to diffuse gas illuminated 
by the ambient ISRF, and have therefore removed all the data 
corresponding to lines of sight toward hot stars.

The TDR models predict column densities of CH$^{+}$
in good agreement with the observations for low densities ($n_{\rm H} \leqslant 100$
cm$^{-3}$). This density limit corresponds to the observed 
average turbulent energy to which we scale our computations
(Eq. \ref{EqNVA}): if ${\overline \varepsilon}_{obs}$ was 
larger, this limit would be larger too.
Most gratifying is the fact that this is achieved without
producing an excess of CH. It is so because, in the vortex, CH is a product
of CH$^{+}$ (its production is therefore enhanced by the turbulent dissipation,
see Appendix \ref{Network}) but its destruction is also enhanced
since it proceeds through endo-energetic reactions (CH + H, $-\Delta E /
k$=2200 K and CH + H$_2$, $-\Delta E / k$=1760 K).
Note that in panels (c,d), because of the assumed homogeneity of the gas
(uniform $A_V$ and $n_{\rm H}$ along the line of sight), the column densities
computed in the TDR models are proportional to $N_{\rm H}$,
hence the straight lines.

The dissipation of turbulent energy is also a plausible explanation for
the excitation of the pure rotational levels of molecular hydrogen:
the TDR models reproduce not only the CH$^{+}$ to H$_2^{*}(J=4)$
column density ratio but also a significant fraction of the large
dynamic range (2 orders of magnitude) over which the correlation is
observed (Fig. \ref{FigCorrelationUV} e,f).  We recall that all the
column densities of species dominated by the warm chemistry are
proportional to $\overline{\varepsilon}_{obs}$ and may be under- or
over-estimated by about a factor of 10.
 
An encouraging result is also shown in Fig. \ref{FigCorrelationUV}
(g,h).  
Gry et al. (2002) and Lacour et al. (2005) found that the H$_2$ excitation
diagrams obtained in the direction of stars later than B2 cannot be reproduced
with the stellar UV excitation only. For comparison we display
a PDR and a TDR model, both computed for $n_{\rm{H}} = 100$ cm$^{-3}$, a value
close to the optimal density found by Nehm\'e et al. (2008b) toward
\object{HD102065}.

The dependence of the TDR models on the density significantly differs 
from those of the PDR models. They share the same dependence 
of the chemistry on gas density but the TDR models have the additional 
dependence on density of the number of vortices and their size. 
The column density of CH$^{+}$  is therefore 
highly dependent on $n_{\rm H}$ (varies roughly
as $n_{\rm H}^{-2}$) while those of CH and H$_2^{*}$($J \geqslant 3$) are almost
insensitive to the density in the range of parameters explored.
Last, the turbulent rate of strain $a$ has a weak effect: all the
column densities decrease by less than a factor of 10 when $a$
increases by a factor of 100.

\subsection{Submillimeter and millimeter observations: absorption lines 
towards continuum sources} \label{ObsRadio}

Conspicuous correlations have been observed in the diffuse ISM among
the observed column densities of OH, H$_2$O, HCO$^{+}$, CO, C$_2$H,
CN, HCN and HNC seen in absorption  along lines of sight towards extragalactic
radio sources or star forming regions.
The column densities of several species appear to be linearly
correlated.  Some are tight correlations such as 
$N({\rm OH})/N(\HCOp) = 27 \pm 7$ (Lucas \& Liszt 1996), $N({\rm CN})/N({\rm
HCN}) = 6.8\pm 1$, and $N({\rm HNC})/N({\rm HCN}) 
= 0.21 \pm 0.05$ (Liszt \& Lucas 2001) in 
the direction of extragalactic radio sources.  Others are looser
correlations, still linear, such as $N(\wat)/N({\HCOp}) \sim 6$ (Olofsson et
al. submitted) and $N(\wat)/N({\rm OH}) \sim 0.3$ (Neufeld et
al. 2002) found in several velocity components towards galactic star forming
regions. The column densities of CO and \CtH\ appear 
non-linearly correlated to those of \HCOp\ (Liszt \& Lucas 1998; 
Lucas \& Liszt 2000). In all cases, these correlations are observed over a
dynamic range of column densities as large as 30-100.

\begin{figure*}[!ht]
\begin{center}
\includegraphics[width=15.5cm,angle=0]{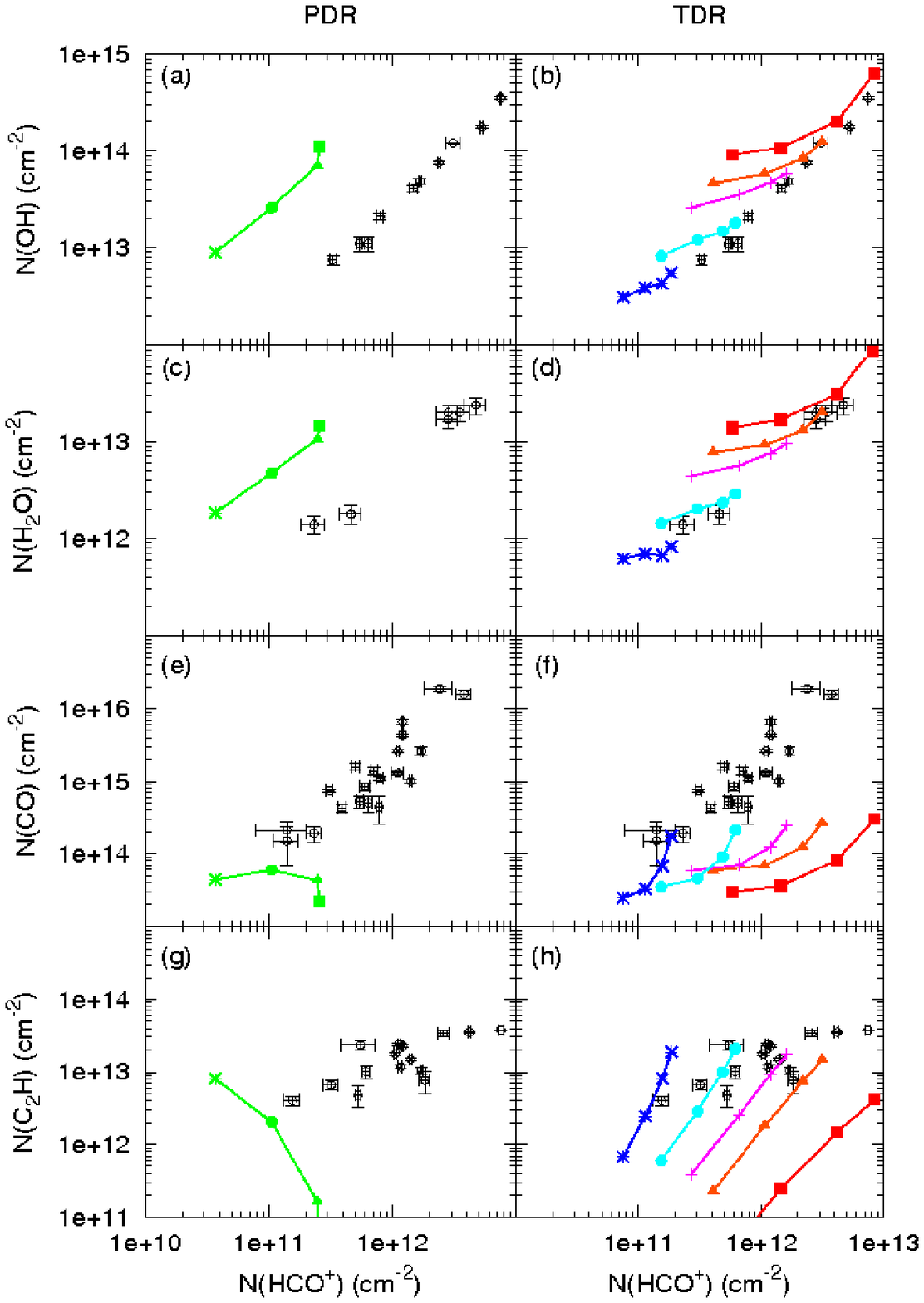}
\caption{Observations compared to PDR models (left panels) and to
  TDR models (right panels). {\bf Data} (open circles) - The data of panels
  (a,b), (c,d), (e,f) and (g,h) are from Lucas \& Liszt (1996), Olofsson et
  al. (submitted to A\&A), Liszt \& Lucas (1998) and Lucas \& Liszt (2000)
  respectively. {\bf PDR and TDR models} (filled symbols) - Computed for
  several densities: 10 (squares), 30 (triangles), 50 (crosses), 100 (circles)
  and 200 (double crosses) cm$^{-3}$. All models are computed for $N_{\rm H} = 1.8
  \times 10^{21}$ cm$^{-2}$ and the HCO$^{+}$ recombination rate of
  Ganguli et al. (1988). The TDR models are
  computed for $A_V=0.4$ and $a$ varying along each curve between $10^{-11}$
  (top right) and $3 \times 10^{-10}$ s$^{-1}$ (bottom left).}
\label{FigCorrelationRadio}
\end{center}
\end{figure*}

\begin{figure*}[!ht]
\begin{center}
\includegraphics[width=15cm,angle=0]{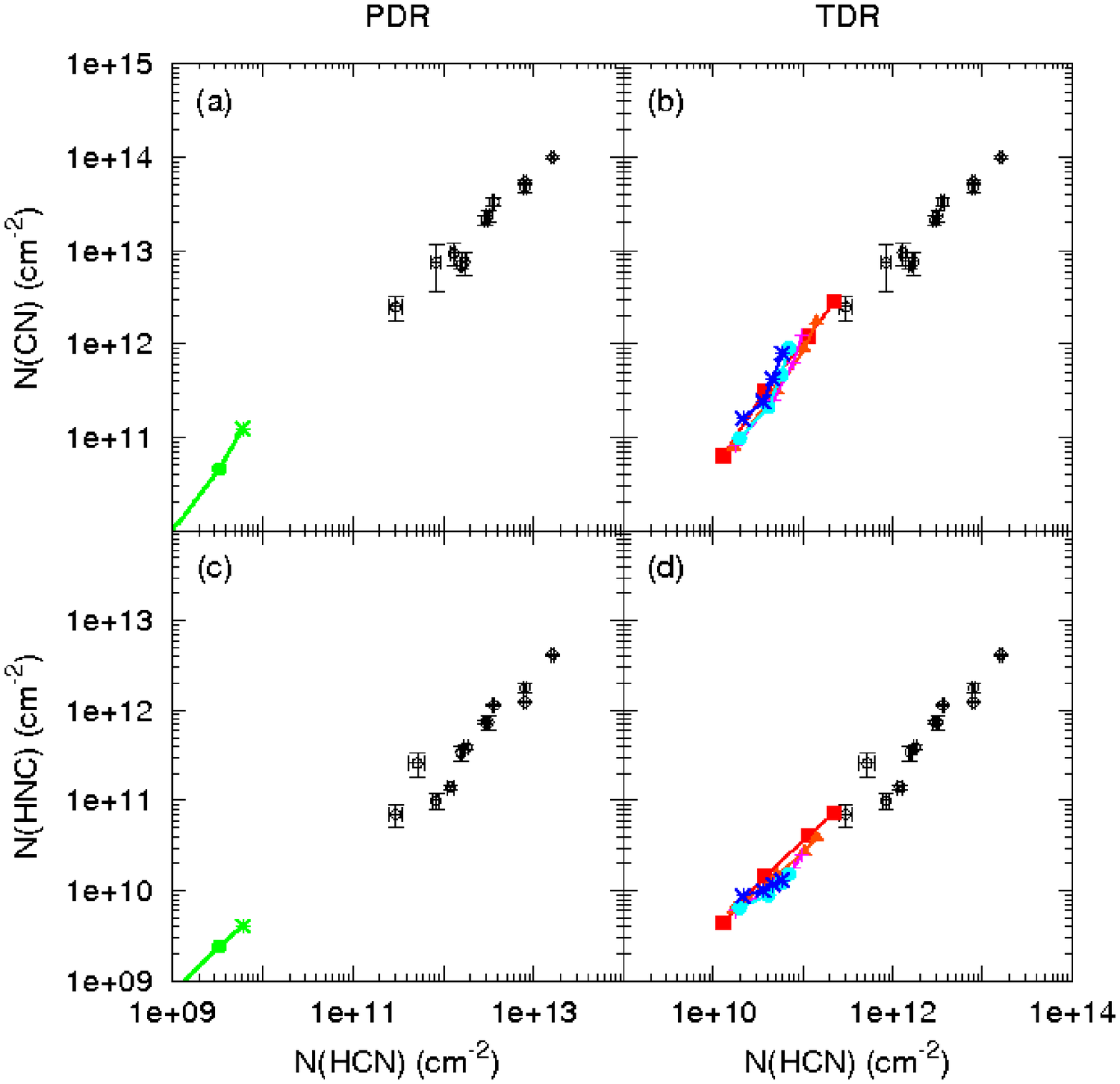}
\caption{Observations compared to PDR models (left panels) and to
  TDR models (right panels). {\bf Data} (open circles) - The data of panels
  (a,b,c,d) are from Liszt \& Lucas (2001). {\bf PDR and TDR models} (filled
  symbols) - Same as Fig. \ref{FigCorrelationRadio}.}
\label{FigCorrelationRadio2}
\end{center}
\end{figure*}

To determine whether these large dynamic ranges correspond to actual variations
of molecular relative abundances or are due to a large range 
of column densities
of gas sampled, we have estimated the total amount of gas $N_{\rm H}= N({\rm
HI})+2N(\HH)$ along the lines of sight used in our study, whenever it
was possible.  Using the $\lambda$21 cm observations of HI (Du Puy et
al. 1969; Radhakrishnan et al. 1972; Lazareff 1975; Dickey et al. 1978)
on the one side, and the measurement of the $\lambda$9 cm line
of CH and the remarkable correlation between CH and H$_2$ in the
diffuse ISM (Liszt \& Lucas 2002) on the other side, we estimated the
total H column density for several lines of sight diplayed in
Fig. \ref{FigCorrelationRadio} and \ref{FigCorrelationRadio2}.  We
found that most lines of sight sample about 1 magnitude of gas, with
the exception of the upper point of panels (a,b) that corresponds to
$2.5 \times 10^{21} < N_{\rm H} < 3.6 \times 10^{21}$ cm$^{-2}$ 
and the lowest for which $N_{\rm H} =
6 \times 10^{20}$ cm$^{-2}$. The next 2 points both verify $N_{\rm H} \geqslant 1.2
\times 10^{21}$ cm$^{-2}$.  The dynamic range of 30-100 observed on the
column densities in Fig. \ref{FigCorrelationRadio} and
\ref{FigCorrelationRadio2} is therefore due in a small part to the amount of
gas on the line of sight, and the actual dynamic range of observed
relative abundances, $N(X)/N_{\rm H}$, remains as large as about 10
along these diffuse lines of sight.

All these data are displayed in Fig. \ref{FigCorrelationRadio} and
\ref{FigCorrelationRadio2}.  The left panels show the predictions of
several PDR models while the right panels display those of the
TDR models for comparison (normalized to 1 magnitude of diffuse
gas sampled on the line of sight).  The calculations  
were made assuming a recombination rate of HCO$^{+}$ of $k
= 2.4 \times 10^{-7} (T/300 \textrm{K})^{-0.69}$ cm$^{-3}$ s$^{-1}$ (Ganguli
1988). This recombination rate being
critical for some of the molecules presented here, 
we discuss the impact of this choice in Appendix \ref{ImpactRecomb}.

\subsubsection{OH, \wat\ and \HCOp}

For these molecules, the PDR models predict linear correlations between 
their abundances, but they fail to reproduce the observed ratios and 
the observed dynamic range of absolute abundances.

The TDR models, without
fine-tuning of the parameters, are consistent with the data over a
broad range of gas densities 10 cm$^{-3}$ $\leqslant n_{\rm H} \leqslant 200$
cm$^{-3}$ and rates of strain 10$^{-11}$ s$^{-1}$ $\leqslant a \leqslant
10^{-10}$ s$^{-1}$. Better agreement is obtained for
small values of the rate of strain, \ie\ a chemistry dominated
by ion-neutral drift heating. Not only are the abundance ratios
  correctly reproduced by the TDR models, 
but the dynamic ranges are also reproduced even with a single value of
$\overline{\varepsilon}_{obs}$. Moreover these results are obtained without
producing  water abundances in excess of observed values. The models at high
densities are consistent with the upper limit reported in HST-GHRS
observations by Spaans et al. (1998).

\subsubsection{C$_2$H and \HCOp}

In the case of C$_2$H, we find that the observed column densities are
reproduced by several models, something that the PDR models cannot even
approach. The average abundance ratio $N($C$_2$H$)/N(\HCOp) = 14.5 \pm 6.7$
(Lucas \& Liszt 2000) is reproduced for almost
the same models that match the observed column densities of OH and H$_2$O: 
30 cm$^{-3}$ $\leqslant n_{\rm H} \leqslant 200$ cm$^{-3}$ and rates of strain
10$^{-11}$ s$^{-1}$ $\leqslant a \leqslant 10^{-10}$ s$^{-1}$.

\subsubsection{CO and \HCOp}

In the case of CO, the TDR models are closer to
the observed column densities than the PDR model predictions, and
again, high densities and low rates of strain are more favorable.
In Appendix \ref{ImpactTauv} we show, however, that if the energy dissipated 
in a vortex lifetime  $\tau_V$ is reduced by a factor of 10 (\ie\ the importance
of the relaxation phase is increased by a factor of 10), the whole range of
observed CO column densities is naturally reproduced  for $n_{\rm H} \geqslant
100$ \cc\ without significantly modifying the results regarding the other
molecules. 

The clumpiness and large fluctuations of density along the line of
sight - often required to find an agreement 
between observed and predicted column densities
in UV-driven chemical models (\eg\ Black \& Dalgarno, 1977; 
Le Petit et al. 2006) - are not required here.

\subsubsection{CN, HCN and HNC}

The TDR models fail to reproduce the high column densities observed for CN,
HCN and HNC, although the results are one to two orders of magnitude above the
PDR models predictions. This discrepancy might be due to the nitrogen
chemistry which involves neutral-neutral reactions whose rates are poorly
known (Pineau des For\^ets et al. 1990, Boger \& Sternberg 2005).

\subsection{The departure of carbon from ionization equilibrium}

\begin{figure}[!ht]
\begin{center}
\includegraphics[width=9cm,angle=0]{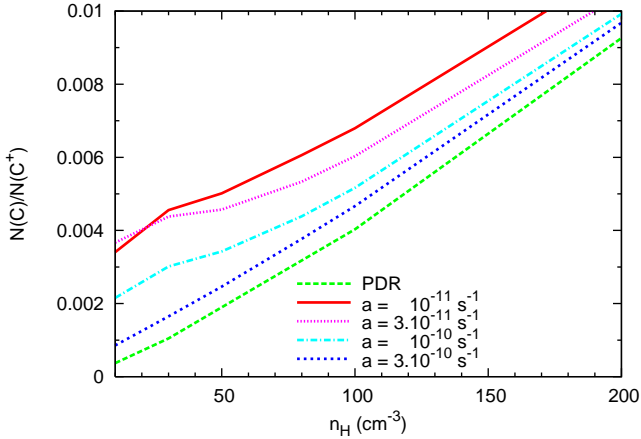}
\caption{Column densities ratio $N($C$)/N($C$^{+})$ computed with the PDR and
  TDR codes, as functions of the density $n_{\rm H}$.}
\label{FigEcartCp}
\end{center}
\end{figure}

Fig. \ref{FigEcartCp} shows that the neutral carbon abundance predicted by
the TDR models is higher by up to a factor of 10 than that of the PDR models
where carbon is in ionization balance.
This may be related to the finding of Fitzpatrick \& Spitzer (1997) 
that the electron density inferred from C and C$^{+}$  is higher than
those inferred from other pairs of ions and neutrals (Mg, S, Ca).

\section{Discussion} \label{Discussion}

\subsection{The gas homogeneity}

In our TDR models, the ambient chemistry is not treated with accurate UV
radiative transfer. Instead, we assume that all the CNM on the line of sight
has the same shielding from the ambient UV radiation field, no matter
its density 
and column density. This takes into account the fractal structure of the CNM
(Elmegreen et al. 2001) also found in the numerical simulations of Audit \&
Hennebelle (2005) and similar to that of edges of molecular clouds (Falgarone
et al. 1991) and translucent molecular clouds (Stutzki et al. 1998). It
also takes into account the fact that fragments that bear molecules are not
isolated in space but are shielded from the ambient UV field by their
neighbouring 
fragments (see Dobbs, Glover \& Klessen 2008). This however does not have a
major impact on our results for all the species that form predominantly in the
warm chemistry, \ie\ most of the species discussed here.

\subsection{The assumption of isochoric evolution} \label{Discuss-Dynamics}

During the vortex stage, the incompressibilty is reasonably justified.
(1) Radially, the
advection force is 3 to $10^{3}$ times larger than the neutral thermal
pressure gradients $\nabla_r \left( n_n k T_n \right)$ over the whole
structure. Since it is also $10^{2}$ to $10^{3}$ times greater than
the  friction force $F_{in}$ (see Sect. \ref{Magnetprop}), $\nabla_r \left(
n_n k T_n \right) \sim 1 - 30 F_{in}$. The ion thermal pressure
gradients $\nabla_r \left( n_i k T_i \right) \sim 2\cdot10^{-4} \nabla_r
\left( n_n k T_n \right)$ are consequently negligible compared to
$F_{in}$. The assumption in Eq. (\ref{EqIons}), involved in the computation of
the steady state, is therefore jusitified. 
(2) Along the $z$ axis $\nabla_z \left( n_n k T_n
  \right)$ is smaller than the advection force as long as $a \geqslant
  10^{-10}$ s$^{-1}$. For $a \leqslant 10^{-10}$ s$^{-1}$ these terms are
  comparable. In this case, the axial expansion timescale is $\sim
  1000$ yr, within a factor of 3 of the vortex lifetime (see
  Fig. \ref{FigVarParam}).

During the relaxation phase, the thermal pressure gradients are dominant and
therefore drive its dynamical evolution. According to the temperature gradients of
Fig. \ref{FigDyn} (panel c), the inner regions ($r \leqslant 0.5 r_0$) should
be compressed while the outer regions ($r \geqslant 1.5 r_0$) should
expand. The characteristic timescales associated with those motions are found to
be short compared to the
chemical timescales (see Sect. \ref{ChemVR}). However, an isochoric relaxation
was assumed because the regions that are chemically the richest and have the
largest contribution to the column densities are those where the thermal
pressure gradients are the smallest ($0.5 r_0 \leqslant r \leqslant 1.5 r_0$).

\subsection{Loosening the vortex framework}

The vortex is a model that has the great and unique advantage of
analytically coupling the large and small scales, but the knowledge we have of
its lifetime is drawn from laboratory experiments only. 
This lifetime $\tau_V$ however plays an important role in the
modelling of a diffuse line of sight: Eq. (\ref{EqColDens}) shows
that if a species is predominantly produced during the relaxation phase,
  its column density is proportional to $1/\tau_V$.

We have made the choice in Sect. \ref{lifetime} to infer $\tau_V$ from the
total energy $E$ dissipated in the vortex, under the constraint
that $\tau_V$ remains larger than the vortex period. We show in Appendix
\ref{ImpactTauv} how the TDR model results are modified when $E$ (and
therefore $\tau_V$) is ten times larger or smaller than the adopted value. As
expected, the impact of $E$ is greater for the molecules that have a long relaxation
time. Fig. \ref{FigVarTauV} shows that a small value of $E$ leads to a better
agreement with the available observations. 
This is shown for OH and CO but the same behavior is observed for CN, HCN
and HNC.
Such a result is in favour of very short bursts of turbulent dissipation, for
which $\tau_V$ may become (at small $a$) shorter than the vortex period.

A non steady state model of a short-lived but intense velocity-shear may thus
be more realistic and produce an equivalent chemical enrichment. This suggests
that the vortex framework is not essential.

\subsection{The chemistry}

One source of uncertainties is undoubtedly linked to the
chemistry itself. While the results are weakly dependent on the
dynamic parameters $a$ and $n_{\rm H}$, a few reaction rates are critical. This is
illustrated by Fig. \ref{FigRadio.Mitchell} that
displays the predictions of the PDR and TDR models for a recent value
(measured in the laboratory, Mitchell \& Mitchell 2006) of the recombination rate
of HCO$^{+}$.

The uncertainties on other chemical rates such as 
the recombination rate of HCNH$^{+}$ (Mitchell \& Mitchell 2006) and the CN
photodissociation rate (Kopp 1996), and neglected reactions (with negative
ions for instance,  see Dalgarno \& MacCray 1973) might affect our modelling.

In this respect, a promising route would be to further link 
the observations of Sect. \ref{ObsUV}
and those of Sect. \ref{ObsRadio} by using the species that are
detectable in both UV and radio domains: for instance CH, CN,
OH and CO. These molecules would permit us to
establish valuable correlations between CH$^{+}$ and the oxygen
bearing species. It would also be a way to evaluate the column density
of molecular hydrogen using two independent methods: the remarkable
correlation between CH and H$_2$ (Liszt \& Lucas 2002) and the less
reliable correlation between CN and H$_2$ (Liszt \& Lucas 2001).

\section{Conclusions and perspectives}

We have built TDR models of diffuse gas in
which the gas temperature and 
chemistry are driven by small-scale bursts of turbulence dissipation. We use
the
framework of a modified Burgers vortex to analytically couple the
large scales of the ambient turbulence to the small scales where
dissipation actually occurs, and to compute the ion-neutral drift
generated by the large neutral accelerations in the vortex.  The
resulting timescales are short and comparable to those of chemical
evolution, which necessitates a non-equilibrium thermal and chemical
approach.

The main feature of these TDR models is that, for the first time, we
quantify the coexistence on a random line of sight across the medium, of
a number of vortices in a stage of active dissipation with
gas in thermal and chemical relaxation, after the end of the
dissipation burst. We also constrain the number of dissipative structures
on a line of sight by the average turbulent energy available in the local
ISM and its transfer rate in the cascade. The
key parameter is the 
turbulent rate of strain $a$ due to the ambient turbulence.

We find that these bursts of dissipation, short-lived and localized,
fill at most a few percent of a random line of sight but
have a  measurable impact on the molecular abundances in the diffuse medium.
For a broad range of
rates of strain and densities, the TDR models 
reproduce the \CHp\ column densities
observed in the diffuse medium and their correlation with highly
excited \HH.  They do so without producing an excess of CH. 

As a natural consequence, they reproduce  
the correct abundance ratios of \HCOp/OH and \HCOp/\wat,  
and the dynamic range of about one order of magnitude over which 
they are observed. Larger C$_2$H and CO abundances than found in other 
types of models, are additional outcomes of the TDR models 
that compare reasonably well with the observed values and their relation to
the \HCOp\ abundances.
Those results are found for a broad range of physical parameters, with
rates of strain in the range $10^{-11}$ s$^{-1}$ $\leqslant a
\leqslant 10^{-10}$ s$^{-1}$.

We find that neutral carbon exceeds the abundance expected at
ionization equilibrium, in agreement with fine-structure line
observations. The abundances and column densities computed for CN, HCN and HNC
are one order of magnitude above PDR predictions and close to the observed
values, although a discrepancy still exists. 

The comparison with observed column densities favors chemical enrichment
dominated by ion-neutral friction, involving shear structures of radius $\sim
100$ AU or more.
It also favors short dissipation bursts. Most of the species are then produced
during the relaxation phase. In this case, their line profile loses the
dynamic signatures of the vortex. The fact that some species are formed in the
active phases and some others in the relaxation phases might explain the
discrepancy among the physical gas parameters inferred from different species
along similar lines of sight.

It is foreseeable that the vortex framework and the assumption
of an isochoric relaxation will have to be superseded by numerical simulations
to: \\
(1) compute the decoupling of magnetic field and neutrals 
in the intense small-scale velocity-shears generated by intermittency, with
boundary conditions imposed by large scale turbulence, \\
(2) take into account the effect of the field on the velocity shears,\\
(3) accurately involve all the pressure gradients (thermal and magnetic)
in the relaxation phase, and \\
(4) take advantge of the huge amount of information contained in the shape of
the line profiles. The difficulty will remain to couple the large and small
scales, over at least 5 orders of magnitude, that is
critical to satisfy the energy requirements of the TDR models and is
currently beyond the numerical capabilities.

\acknowledgement
We are most grateful to John Black for pointing out the uncertainties
concerning the \HCOp\ recombination rate and to Brian Mitchell for the
detailed informations regarding the past and most recent laboratory
measurements of this reaction. We thank Javier Goicoechea for the valuable
discussions on chemistry in UV-dominated environments. 
We also thank the referee for the thorough and encouraging comments on this
  manuscript.
\endacknowledgement

\appendix
\section{The recombination of HCO$^{+}$} \label{ImpactRecomb}

\begin{figure*}[!ht]
\begin{center}
\includegraphics[width=15.5cm,angle=0]{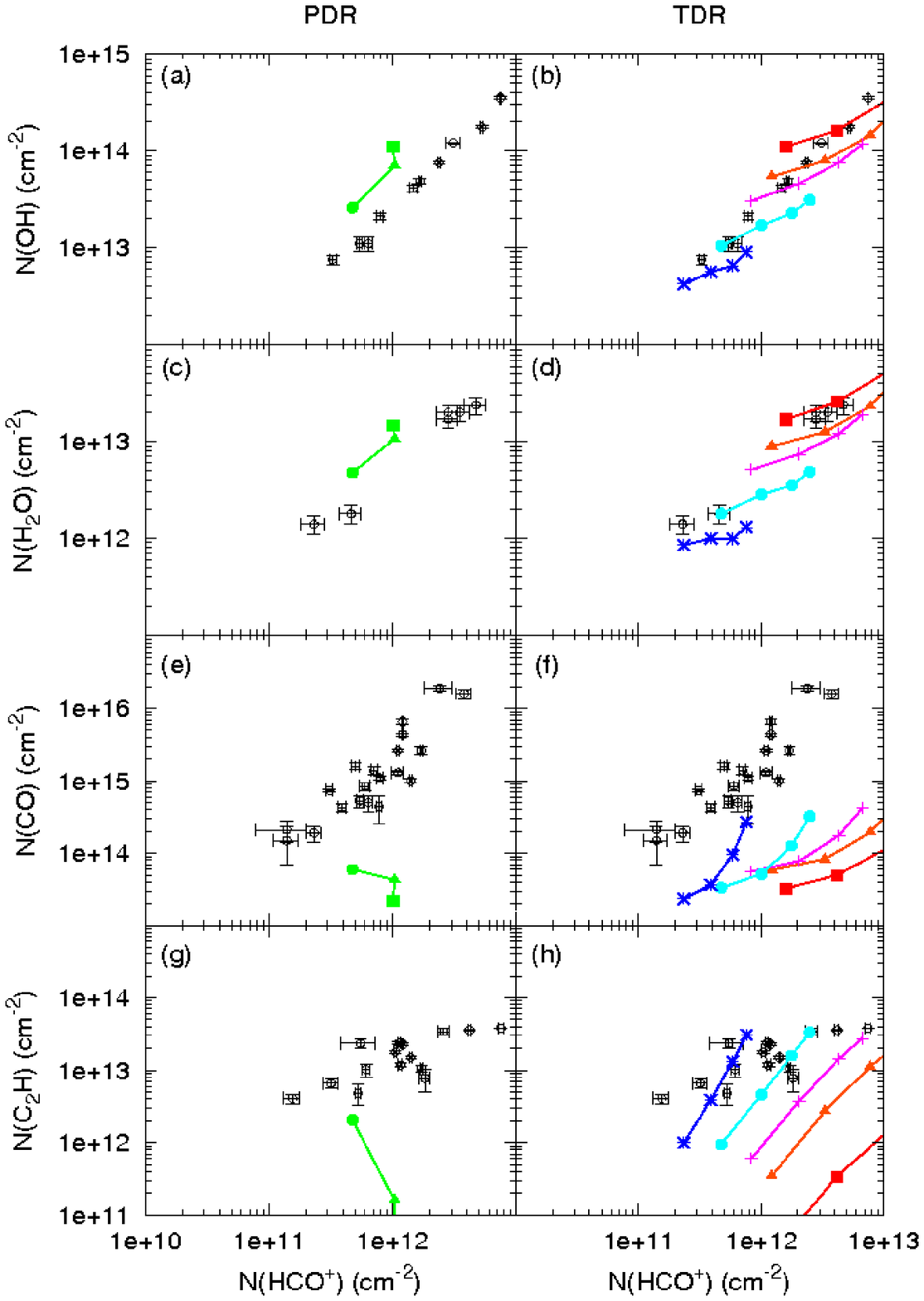}
\caption{Observations compared to PDR models (left panels) and to
  TDR models (right panels). {\bf Data} (open circles) - Same as
  Fig. \ref{FigCorrelationRadio}. {\bf PDR and TDR models} (filled symbols) -
  Same as Fig. \ref{FigCorrelationRadio} except for the recombination rate of
  HCO$^{+}$ chosen to be $0.7 \times 10^{-7} (T/300 \textrm{ K})^{-0.5}$
  cm$^{3}$ s$^{-1}$ (Mitchell \& Mitchell 2006).}
\label{FigRadio.Mitchell}
\end{center}
\end{figure*}

One critical reaction rate of our work is the recombination of
HCO$^{+}$ which determines the equilibrium between the oxygen bearing
molecules (OH, CO, H$_2$O, ...). The associated rate  $k$
has varied over more than an order of magnitude during the last 30
years among various experiments.

Adams et al. (1982) and Amano (1990) performed measurements of $k$
(studying respectively a flowing and a stationary afterglow plasma) in a
temperature range of  $100 \textrm{K} \leqslant T \leqslant 300 \textrm{K}$
and found respectively $1.1 \times 10^{-7}$  cm$^{3}$ s$^{-1}$ at 300 K and
$3.1 \times 10^{-7}$ cm$^{3}$ s$^{-1}$ at 273 K. In 1988 Ganguli et
al. (stationary afterglow plasma) explored the high energy domain $293
\textrm{K} \leqslant T \leqslant 5500 \textrm{K}$ and obtained a scaling law:
\begin{equation}
 k = 2.4 \times 10^{-7} \times (T/300 \textrm{K})^{-0.69} \quad \textrm{cm}^{3}\textrm{s}^{-1}.
\end{equation}
In approximatively the same range of energy (but with the merged beam
measurement technique) Le Padellec et al. (1997) obtained 
\begin{equation}
k = 1.7 \times 10^{-7}(T/300 \textrm{K})^{-1.2} \quad \textrm{cm}^{3}\textrm{s}^{-1}.
\end{equation}
Later the same group (Mitchell \& Mitchell 2006) revised their previous value by
taking into account the exploration of the low energy domain ($\leqslant 0.01$
eV). They found :
\begin{equation}
k = 0.7 \times 10^{-7} \times (T/300 \textrm{K})^{-0.5} \quad \textrm{cm}^{3}\textrm{s}^{-1}.
\end{equation}
More recently Korolov et al. (in press, flowing aftergow plasma) performed
measurements in the range $150 \textrm{K} \leqslant T \leqslant 270
\textrm{K}$. They obtained results in good agreement with those of Le Padellec
et al. (1997), this is to say a steep dependence on the temperature:
\begin{equation}
k = 2.0 \times 10^{-7} \times (T/300 \textrm{K})^{-1.3} \quad \textrm{cm}^{3}\textrm{s}^{-1}.
\end{equation}

Fig. \ref{FigRadio.Mitchell} displays the results we obtain with the TDR
and PDR models assuming the rate of Mitchell \& Mitchell (2006) which is
the smallest yet found. In comparison Fig. \ref{FigCorrelationRadio} was
obtained assuming the rate of Ganguli et al. (1988) which is one of the
highest and is usually adopted in other chemical networks (UMIST
database; OSU database).

\section{Influence of the total energy $E$ dissipated per vortex. } \label{ImpactTauv}

\begin{figure*}[!ht]
\begin{center}
\includegraphics[width=14.8cm,angle=0]{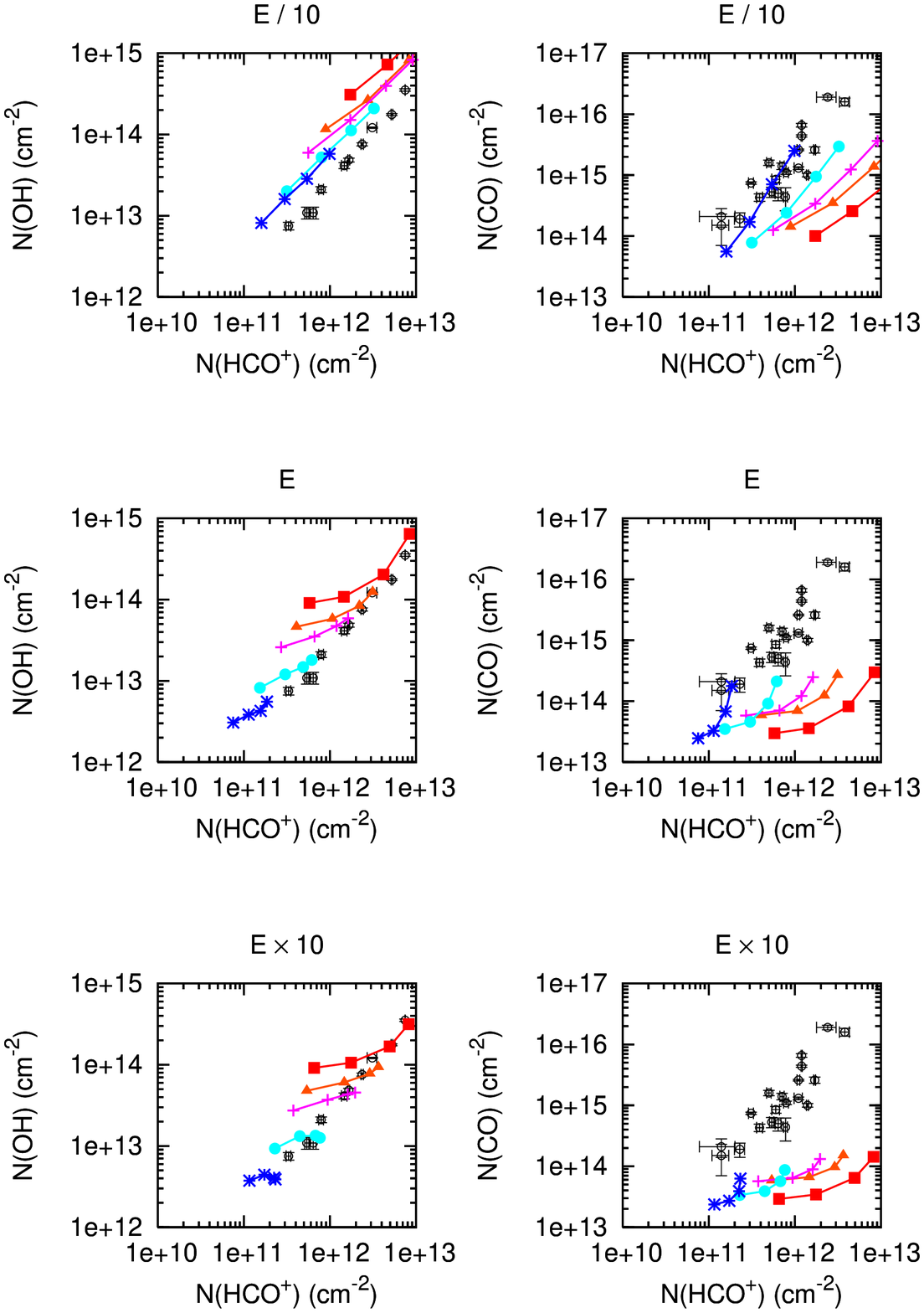}
\caption{Observations compared to the TDR models. The data are from Lucas \&
  Liszt (1996) and Liszt \& Lucas (1998). The TDR models are scaled to $N_{\rm
  H} = 1.8 \times 10^{21}$ cm$^{-2}$ and are computed for several values of
  the total energy $E$ dissipated by a vortex: from 0.1 (top panels) to 10
  times (bottom panels) that of the reference model. The symbols are the same as
  in Fig. \ref{FigCorrelationRadio}.}
\label{FigVarTauV}
\end{center}
\end{figure*}

  Fig. \ref{FigVarTauV} illustrates the effect on the column densities of OH,
  CO and HCO$^{+}$ of varying the energy dissipated in the vortex $E$ by
  two orders of magnitude.

  Fig. \ref{FigVarTauV} suggests that small values of $E$ lead to better
  agreement with the observations. Smaller amounts of dissipated energy, in
  our framework, means shorter vortex lifetimes, that accordingly give a
  larger relative importance to the relaxation phase (Eq. \ref{EqColDens}). 
  In the models displayed
  on the top panels, the lifetime is such that the relaxation phase
  dominates the production of almost all the species we are interested in,
  namely CH, C$_2$H, OH, H$_2$O, HCO$^{+}$, CO, CN, HCN and HNC.

\section{UV-dominated versus turbulence-dominated chemical networks} \label{Network}

\begin{figure*}[!ht]
\begin{center}
\includegraphics[width=17cm,angle=0]{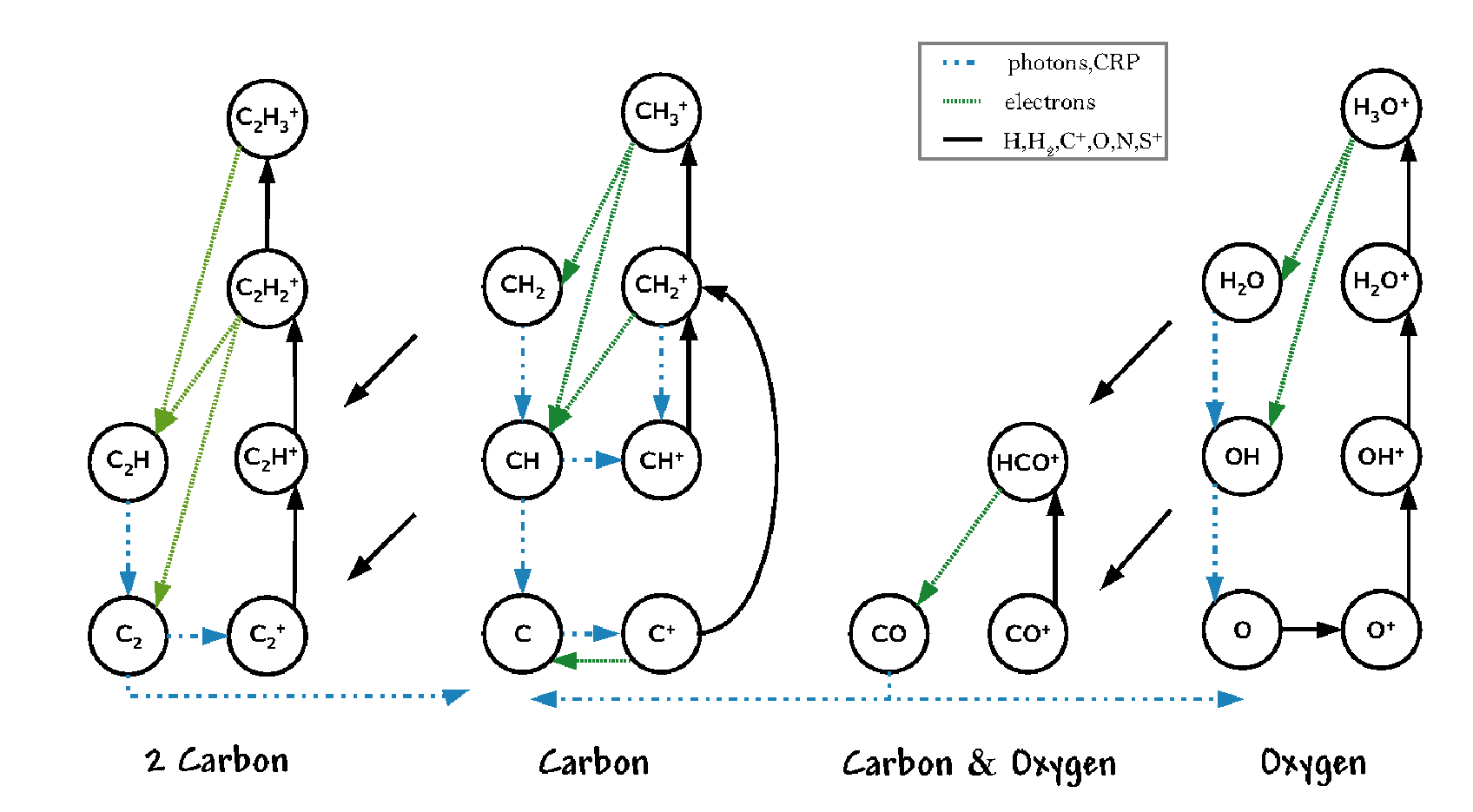}
\caption{Chemical network of a UV-dominated chemistry: $n_{\rm H} = 30$
  cm$^{-3}$ and $A_V$ = 0.4. This figure is simplified: for each species
  only the dominant reaction of production and the dominant reaction of
  destruction are displayed.}
\label{ChemNetwork1}
\end{center}
\end{figure*}

\begin{figure*}[!ht]
\begin{center}
\includegraphics[width=17cm,angle=0]{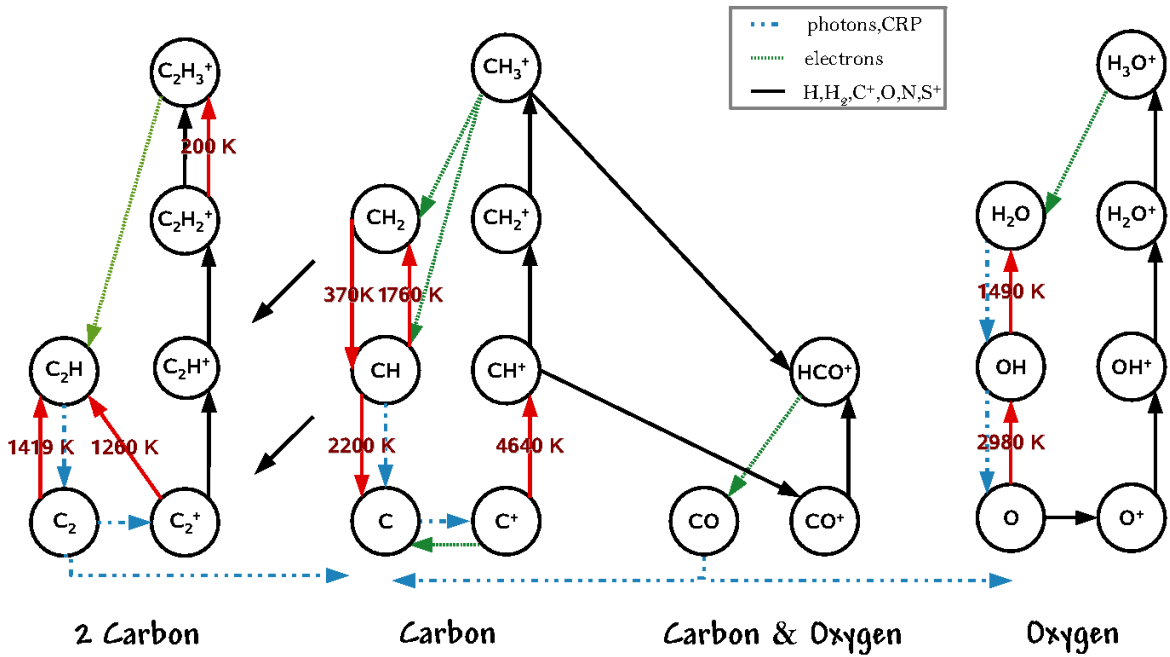}
\caption{Same as Fig. \ref{ChemNetwork1} for a turbulence-dominated chemistry:
  $n_{\rm H} = 30$ cm$^{-3}$, $A_V$ = 0.4, $a=3\times10^{-11}$ s$^{-1}$ at a
  radius $r=r_0$. The red arrows show the endoenergetic reactions with the
  energy involved.}
\label{ChemNetwork2}
\end{center}
\end{figure*}

  Figs. \ref{ChemNetwork1} and \ref{ChemNetwork2} display the main production
  and destruction routes of the molecules of interest in the ambient
  UV-dominated diffuse medium and in the vortex for the reference TDR model at
  a radius $r=r_0$ respectively. 

  These figures are simplified: \emph{only} the dominant reaction of
  production and the dominant reaction of destruction of the species are
  shown. For example one would expect the reaction O +
  H$_3^{+}$ $\rightarrow$ OH$^{+}$ + H$_2$ to be displayed in
  Fig. \ref{ChemNetwork1}. But in the model presented here, the efficiency of
  this reaction on the production of OH$^{+}$ is only 12\%, while the efficiency
  of O$^{+}$ + H$_2$ $\rightarrow$ OH$^{+}$ + H is 86\%.
  Therefore it does not appear on the diagram.

\end{document}